\documentclass[lettersize,journal]{IEEEtran}
\usepackage{amsmath,amsfonts}
\usepackage{algorithmic}
\usepackage[ruled,linesnumbered,lined]{algorithm2e}
\usepackage{array}
\usepackage[caption=false,font=normalsize,labelfont=sf,textfont=sf]{subfig}
\usepackage{textcomp}
\usepackage{stfloats}
\usepackage{xcolor }
\usepackage{url}
\usepackage{verbatim}
\usepackage{graphicx}
\usepackage{cite}
\begin{document}

\title{Bayesian and Multi-Armed Contextual Meta-Optimization for Efficient Wireless \\Radio Resource Management}

\author{Yunchuan Zhang, \IEEEmembership{Student Member,~IEEE}, Osvaldo Simeone, \IEEEmembership{Fellow,~IEEE}, Sharu Theresa Jose, \IEEEmembership{Member,~IEEE}, Lorenzo Maggi and~Alvaro Valcarce, \IEEEmembership{Senior Member,~IEEE}
\thanks{The work of Osvaldo Simeone was
supported by the European Research Council (ERC) under the European Union’s Horizon 2020 Research and Innovation Programme (grant agreement No. 725732), by the European Union's Horizon Europe project CENTRIC (101096379),  and by an Open Fellowship of the EPSRC (EP/W024101/1).}
\thanks{Yunchuan Zhang and Osvaldo Simeone are with the Department of Engineering, King's College London, London WC2R 2LS, UK ({yunchuan.zhang, osvaldo.simeone}@kcl.ac.uk)}

\thanks{Sharu Theresa Jose is with the Department of Computer Science, University of Birmingham, Birmingham B15 2TT, UK (s.t.jose@bham.ac.uk)}

\thanks{Lorenzo Maggi and Alvaro Valcarce are with Nokia Bell Labs France, Route de Villejust, 91620 Nozay, France ({lorenzo.maggi, alvaro.valcarce$\_$rial}@nokia-bell-labs.com)}
}



\maketitle

\begin{abstract}
Optimal resource allocation in modern communication networks calls for the optimization of objective functions that are only accessible via costly separate evaluations for each candidate solution. The conventional approach carries out the optimization of resource-allocation parameters for each system configuration, characterized, e.g., by topology and traffic statistics, using global search methods such as Bayesian optimization (BO). These methods tend to require a large number of iterations, and hence a large number of key performance indicator (KPI) evaluations.  In this paper, we  propose the use of meta-learning to transfer knowledge from data collected from related, but distinct, configurations in order to speed up optimization on new network configurations. Specifically, we combine meta-learning with BO, as well as with multi-armed bandit (MAB) optimization, with the latter having the potential advantage of operating directly on a discrete search space. Furthermore, we introduce novel contextual meta-BO and meta-MAB algorithms, in which transfer of knowledge across configurations occurs at the level of a mapping from graph-based contextual information to resource-allocation parameters. Experiments for the problem of open loop power control (OLPC) parameter optimization for the uplink of multi-cell multi-antenna systems provide insights into the potential benefits of meta-learning and contextual optimization.
\end{abstract}


\section{Introduction}
\subsection{Context and Scope}
\IEEEPARstart{T}{he} management and configuration of modern cellular communication systems requires the optimization of a large number of parameters that define the operation across all segments of the network, including the radio access network (RAN) \cite{polese2022understanding}. Machine learning, or artificial intelligence (AI), methods are often invoked as potential solutions, and most efforts in this direction leverage neural network-based methods, which may incorporate contextual information such as on the network topology \cite{simeone2018very, he2021overview,yaohua2019survey}. However, the implementation of AI solutions for resource allocation is practically constrained by the limited access of the designer to relevant data and to efficiently computable objective functions. In fact, typically, each candidate solution can only be evaluated via a point-wise estimate of \emph{key performance indicators (KPIs)} through expensive simulations or measurements \cite{Nei2020challenges}. This paper investigates methods that aim at reducing the number of KPI evaluations needed for AI-based resource allocation via the introduction of novel optimizers based on meta-learning \cite{Lisha2022,simeone2020learning}, multi-armed bandit optimization \cite{gai2012combinatorial},  and contextual optimization \cite{krause2011contextual}.

To exemplify the application of the proposed resource-allocation optimizers, we focus on the important problem of \emph{open loop power control (OLPC)} for the uplink of a multi-cell system with multi-antenna base stations \cite{Maggi2021bogp} (see Fig. 1). This optimization requires a search over a large discrete space of candidate options, and each candidate power control parameter set needs to be evaluated via the use of a network simulator or via measurements in the field. The conventional approach carries out the optimization of resource-allocation parameters for each system configuration, which is characterized, e.g., by topology and traffic statistics \cite{yongshun2016power}. This \emph{per-configuration} approach is justified by the diversity of network deployments, which generally prevents the direct reuse of solutions found for one deployment to another deployment. However, as mentioned, this class of solutions is practically impaired by the need to evaluate many candidate solutions as intermediate steps towards a satisfactory solution.

\subsection{Related Work}

Machine learning solutions based on deep neural networks (DNNs) train a generic dense neural network in a supervised or unsupervised fashion to approximate the output of model-based power control algorithms such as the Weighted Minimum Mean Squared Error (WMMSE) \cite{haoran2018tsp,wei2020icassp,fei2020tcomm,woongsup2019wcl,ahmed2019network,ashok2022tccn}. Alternatively, reinforcement learning can be leveraged to autonomously optimize channel
selection and power allocation based on feedback from the network designer  \cite{junjie2021twc}. Unlike methods based on supervised or unsupervised learning,  reinforcement learning does not rely on a model-based optimizer and it does not require access to gradients of the objective function, but it typically necessitates many evaluations of the KPIs of interest at intermediate solutions. 
\begin{figure*}[t]

  \centering
	
  \includegraphics[scale=0.38]{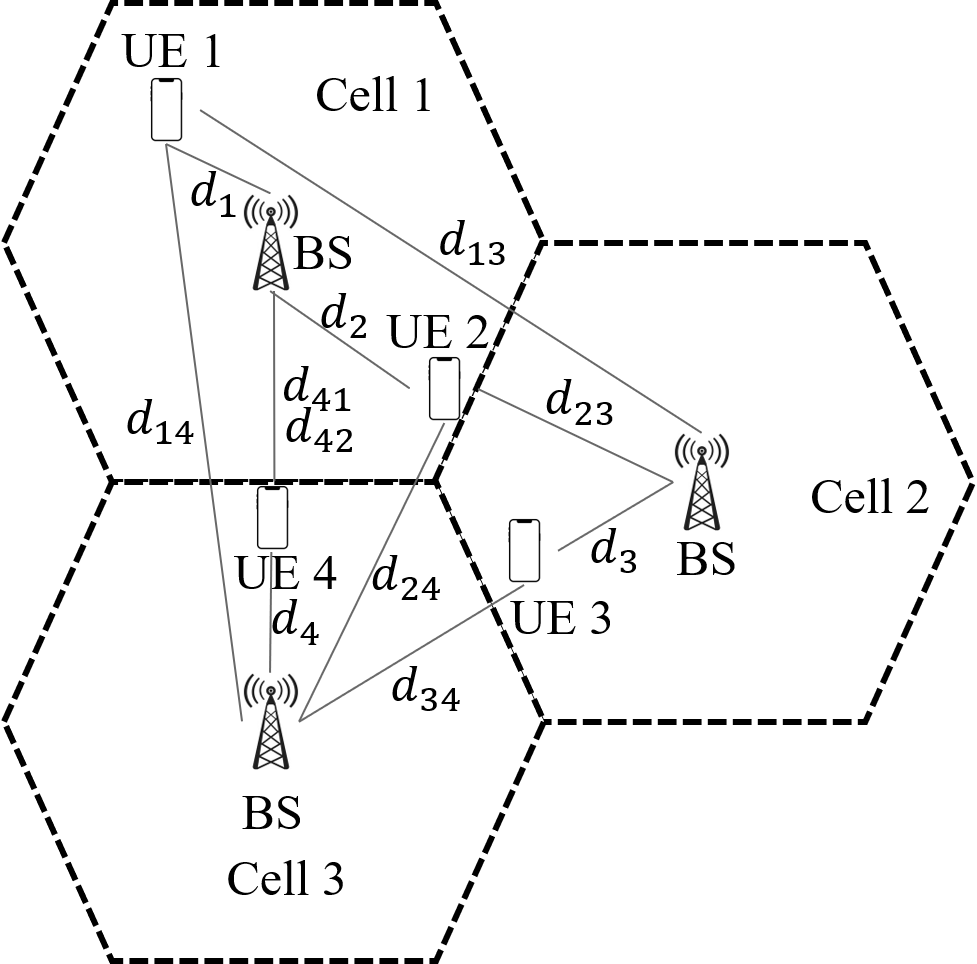}
  \includegraphics[scale=0.55]{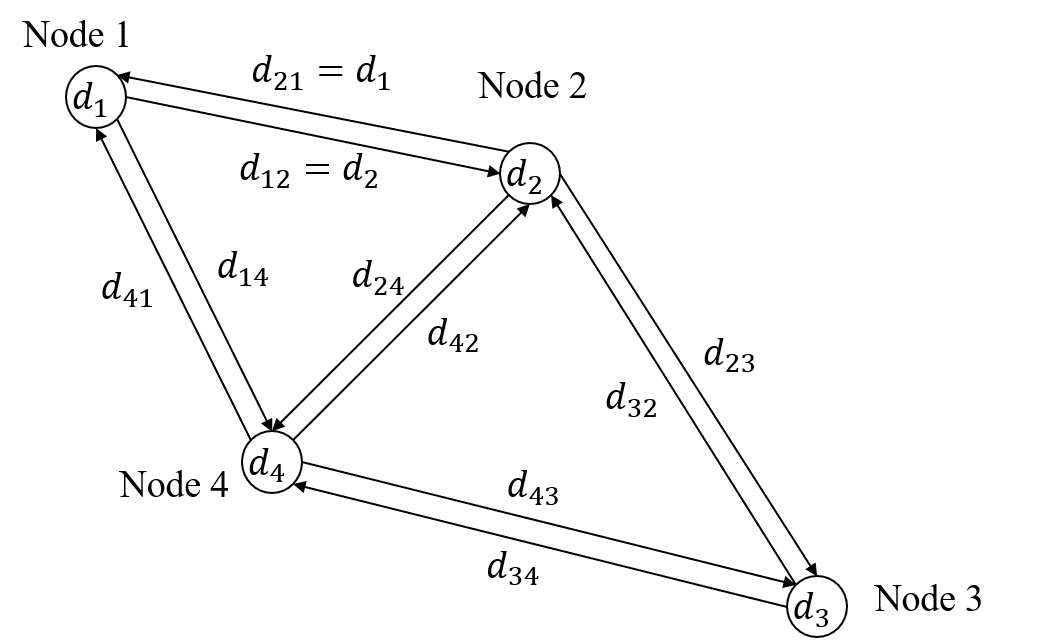}
  \caption{A configuration $\tau$ is described in this example by the network topology illustrated on the left. The network encompasses $N^C=3$ cells, each with one BS. There are four UEs, with $N_{U,1}=2$,  $N_{U,2}=1$, and  $N_{U,3}=1$ UEs in cells 1, 2, and 3, respectively. Therefore, communication links exist between UE 1 and the BS in cell 1, UE 2 and the BS in cell 1, UE 3 and the BS in cell 2, as well as UE 4 and the BS in cell 3. Meta-learning schemes based on BO or MAB  optimize power allocation for this network configuration based on KPI measurement from other network configurations, characterized, e.g., by different distances or number of UEs per cell. Furthermore, as explained in Sec. VI, for contextual optimization, the context vector $\mathbf{c}_\tau$ may contain all distances, where $d_i$ is the distance from UE-$i$ to the serving BS and $d_{ij}$ is the distance between UE-$i$ and the BS serving UE-$j$. The context vector $\mathbf{c}_\tau$ can be described in terms of the interference graph $\mathcal{G}_{\tau}$ shown on the right. In the graph, each node corresponds to one of the four links, and is marked with the relevant distance between UE and serving BS. A directed edge is included between links for which the distance between the transmitting UE for the first link and the receiving BS for the second link is sufficiently small, indicating a meaningful level of interference between the first link and the second link.}
  \label{fig: context graph}
%
\end{figure*}

It was recently pointed out by some of the authors of the present contribution in \cite{Maggi2021bogp} that \emph{Bayesian optimization} (BO) with Gaussian Process (GP) can provide a more flexible solution that does not require access to gradient information for the objective function and can potentially reduce convergence time for power control optimization as compared to reinforcement learning. However, BO still requires a separate optimization for each network configuration, and the number of per-configuration KPI evaluations may  still be prohibitively high.

\emph{Meta-learning}, or learning to learn, is a general paradigm for the design of machine learning algorithms that can transfer knowledge from data related to different tasks, to \emph{any} new, related, task. Knowledge is transferred in the form of an optimized inductive bias that can be realized via a prior over the weights of neural networks \cite{amit2018meta}, an initialization of gradient descent \cite{finn2017model}, or an embedding space shared across auxiliary tasks \cite{NIPS2016_90e13578}, among other solutions. Meta-learning is markedly distinct to other knowledge-transferring paradigms such as transfer learning. In fact, transfer learning focuses on the optimization of a model for a specific target task given data from a given source task. In contrast,  meta-learning optimizes an adaptation procedure -- representing an inductive bias -- that can be applied to any, a priori unknown, related task \cite{simeone2022mle}. 

Applications of meta-learning to communication systems are currently limited to DNN-based models, and encompass demodulation \cite{park2020learning,cohen2022bayesian}, channel prediction \cite{sangwoo2022meta}, beamforming \cite{yuan2020transfer}, feedback design \cite{liu2022learning}, and power control via graph neural networks \cite{nikoloska2022modular}. We refer to \cite{Lisha2022} for an extensive review. As shown recently in \cite{rothfuss2022metabo,ivana2021meta}, meta-learning can be combined with BO to achieve convergence and safe exploration within a smaller number of iterations. Applications of this methodology to resource allocation have yet to be explored.

Contextual BO was studied in \cite{krause2011contextual}. In this reference, the BO optimizer is given a different context vector at each optimization step. For this situation, the authors of \cite{krause2011contextual} propose to append the context vector to the input. This approach does not work well for the problem of interest in which the context vector is fixed at run time, and hence different solutions must be compared for the same context vector. This calls for the use of a distinct context-based optimization approach, which we introduce in this work.

\subsection{Main Contributions}
In this paper, we propose for the first time the use of meta-learning to transfer knowledge from data collected from related, but distinct, network configurations in order to speed up optimization of resource allocation parameters on new network configurations. The speed-up is measured in terms of the number of evaluations of KPIs for candidate solutions that are needed to attain an effective resource allocation strategy. To this end, our contributions are of both methodological and application-based nature. Specifically, we introduce new meta-learning-based design methodologies, which we expect to be of independent interest and broader applicability; and we investigate their application to uplink OLPC in cellular systems. The proposed methods leverage the availability of offline data from multiple network configurations, or deployments, to tailor OLPC adaptation strategies for any new deployment.


The main contributions of the paper are as follows:

\begin{itemize}
    \item[$\bullet$] At a methodological level, we introduce a novel scheme that combines meta-learning with \emph{multi-armed bandit (MAB) optimization} \cite{gai2012combinatorial}. MAB has the potential advantage over BO of operating directly on a discrete search space. This is a particularly useful feature in problems, such as OLPC,  in which the optimization variables are quantized. Our approach, termed \emph{meta-MAB}, is based on a specific parameterization of the Exp3 bandit selection policy \cite{mackay2003} that enables meta-optimization based on data from multiple tasks. 

    \item[$\bullet$] Also at a methodological level, we propose novel \emph{contextual} meta-BO and meta-MAB algorithms that can incorporate task-specific information in the form of a graph. The proposed approach is based on a graph kernel formulation \cite{deepgraphkernel}, whereby problems characterized by similar contextual graph information are assigned related solutions. In the context of the OLPC problem, contextual meta-BO and meta-MAP optimize a mapping from graph-based contextual information about the network topology to power allocation parameters (see Fig. \ref{fig: context graph}).

    \item[$\bullet$] In terms of applications,  we propose for the first time to leverage meta-BO and meta-MAB for optimal resource allocation with a focus on the problem of OLPC parameter optimization. As mentioned, while meta-BO is directly applicable to continuous search spaces, and can also be adapted to work for discrete optimization, meta-MAB directly targets discrete search spaces. The benefit of the proposed meta-BO and meta-MAB strategies is the reduction in the number of KPI evaluations, or iterations, needed to optimize resource allocation for each new configuration.

    \item[$\bullet$] We validate the performance of all the proposed methods in a multi-cell system following 3rd Generation Partnership Project (3GPP) specifications.  Experiments for the problem of OLPC parameter optimization provide insights into the potential benefits of meta-learning and contextual optimization strategies.

\end{itemize}

The rest of the paper is organized as follows. First, in Sec. II we formulate the problem. Sec. III reviews BO and Sec. IV introduces meta-BO; while Sec. V reviews MAB and proposed meta-MAB. Contextual meta-BO and meta-MAB are introduced in Sec. VI, and experimental results are provided in Sec. VII. Sec. VIII concludes the paper.

\section{Problem Formulation}
\label{sec: problem formulation}
We consider the problem of uplink power allocation in a wireless cellular communication system with $N_C$ cells, with each $c$th cell containing one multi-antenna base station (BS) and $N_{U,c}$ user equipments (UEs). As in \cite{Maggi2021bogp}, we specifically focus on the optimization of long-term uplink power control parameters that are network-controlled and updated  infrequently by the network operator. Accordingly, the power-control parameters are not adapted in real time, i.e.,  at time scale of milliseconds, but rather at the scale of hours -- e.g., peak vs. non-peak times -- or days -- e.g., weekday vs. week-end.

In each cell $c$, the BS is equipped with $N_{R,c}$ receiving antennas, and each UE $u$ has $N_{T,c,u}$ transmit antennas. Note that different UEs, such as smart watches, smart phones, or sensors, generally have a distinct number of antennas, which may not be known at the network side. Let $\mathrm{P}_{\mathbf{H}}$ denote the probability distribution of the instantaneous channel state information (CSI) $\mathbf{H}$ describing the propagation channels between the BSs and all the UEs. The channel distribution $\mathrm{P}_{\mathbf{H}}$ may account for the environment type, e.g., rural, urban, or industrial; for the locations of the UEs and BSs; as well as for slow and fast fading effects, including blockages. The user activity can be also implicitly modelled by the distribution $\mathrm{P}_{\mathbf{H}}$, as inactive UEs can be modelled as having negligible connectivity to all BSs. 

We define the \textit{configuration} $\tau$ of the system via the tuple $\tau=(\mathbf{N}_R, \mathbf{N}_U, \mathbf{N}_T, \mathrm{P}_{\mathbf{H}})$ consisting of vectors $\mathbf{N}_R$ and $\mathbf{N}_T$, which collect the numbers of antennas at BSs and UEs across the cells, respectively; of vector $\mathbf{N}_U$, which counts the number of UEs in each cell, and of the CSI distribution $\mathrm{P}_{\mathbf{H}}$. We are interested in developing efficient solutions for power allocation of the UEs given any system configuration $\tau$. We first focus on developing efficient solutions for power allocation of the UEs given any system configuration $\tau$. Then, in Sec. \ref{sec: context}, we consider a more general setting in which the power control policy can also depend on “context” information about the CSI distribution $\mathrm{P}_{\mathbf{H}}$, such as the topology of the network.  

For a given configuration $\tau$, the distribution $\mathrm{P}_{\mathbf{H}}$ is generally unknown. For instance, the UE distribution and/or fading models may not be available. Power control can be based only on the vectors $\mathbf{N}_R, \mathbf{N}_U, \mathbf{N}_T$, as well as on a dataset $\mathcal{D}_{\tau}=\{\mathbf{H}_{\tau,s}\}_{s=1}^{S_{\tau}}$ of $S_{\tau}$ CSI realizations. The dataset $\mathcal{D}_{\tau}$ is practically obtained through channel estimation procedures. Our goal is to design mechanisms that can optimize the power allocation strategy for any new configuration $\tau$ even when only few data points are available, i.e., when $S_{\tau}$ is small, and/or when limited time and computational power can be expended for optimization. To this end, we will combine an offline meta-optimization step with an adaptation step based on dataset $\mathcal{D}_{\tau}$. In practice, as we will discuss, one may not have access to CSI, but only to point-wise measurements of a relevant \textit{key performance indicator (KPI)}, and the aim is to minimize the number of such measurements required to identify a well performing power control solution.

According to the 3GPP's fractional power control policy \cite{ubeda2008fpc}, each UE $u$ in cell $c$ calculates its transmitting power $P^{\mathrm{TX}}_{c,u}$ (in dBm) on the physical uplink shared channel (PUSCH) as a function of the \textit{open loop power control parameters} (OLPC) $(P_{0,c}, \alpha_c)$. These consist of the expected power $P_{0,c}$ received at the BS of cell $c$ under full power compensation, and the fractional power control compensation parameter $\alpha_c\in [0,1]$ for cell $c$. Specifically, focusing on a single resource block, the power $P^{\mathrm{TX}}_{c,u}$ is obtained as \cite{ubeda2008fpc}
\begin{align}
    P^{\mathrm{TX}}_{c,u}=\min\{P^{\mathrm{max}}_{c,u}, P_{0,c}+\alpha_c \mathrm{PL}_{c,u}+\mathrm{CL}_{c,u}\}\quad [\text{dBm}],\label{eq: PUSCH power}
\end{align}
where $P^{\mathrm{max}}_{c,u}$ is the maximum UE transmit power; and $\mathrm{PL}_{c,u}$ is the pathloss in dB towards the serving $c$th BS, and $\mathrm{CL}_{c,u}$ is the closed-loop power control adjustment for UE $u$. Note that, by \eqref{eq: PUSCH power}, if $\alpha_c=1$ the received power is $P_{0,c}+\mathrm{CL}_{c,u}$, unless the maximum power constraint $P^{\mathrm{max}}_{c,u}$ forces the equality $P^{\mathrm{TX}}_{c,u}=P^{\mathrm{max}}_{c,u}$ in \eqref{eq: PUSCH power}. The OLPC parameters ($P_{0,c}, \alpha_c$) are generally distinct across the cells, i.e., they depend on the cell index $c$. Furthermore,  they are constrained to lie in the set of $N_{OLPC}=912$ options described in
Table \ref{table: olpc choices} \cite{Maggi2021bogp}. We define as $\mathbf{P}_{0}=[P_{0,1},\ldots,P_{0,N_C}]^{T}$ the $N_C\times 1$ vector of expected received power parameters across all cells; and as $\boldsymbol{\alpha}=[\alpha_1,...,\alpha_{N_C}]^{T}$ as the vector of fractional power compensation parameters. Note that the optimization space, i.e., the number of allowed values of the OLPC parameters $P_0$ and $\alpha$ grows exponentially with the number of cells $N_C$.
\begin{table}[h!]
\centering
\caption{Allowed values for OLPC parameters}
\label{table: olpc choices}
\begin{tabular}{@{}|c|c|@{}}
\hline
 $P_{0}\; (\mathrm{dBm})$ & $-202, -200, ..., +22, +24$  \\ \hline
 $\alpha$ & $0, 0.4, 0.5, 0.6, 0.7, 0.8, 0.9, 1.0$  \\ \hline
\end{tabular}
\end{table}

The OLPC parameters $(\mathbf{P}_0, \boldsymbol{\alpha})$ are to be selected so as to optimize a given uplink KPI \cite{Maggi2021bogp}. The KPI obtained for a given CSI $\mathbf{H}_{\tau}$ is a function of the OLPC parameters $(\mathbf{P}_{0}, \boldsymbol{\alpha})$ through \eqref{eq: PUSCH power}, and is denoted as $\text{KPI}(\mathbf{P}_0,\boldsymbol{\alpha},\mathbf{H}_{\tau})$. The KPI may be obtained via fixed measurements or through the use of a simulator. For any given configuration $\tau$, we are interested in maximizing the average network-wide KPI as per the discrete optimization problem
\begin{align}
    \max \limits_{\mathbf{P_0},\boldsymbol{\alpha}}\bigg\{\mathbb{E}_{\mathrm{P}_{\mathbf{H}_{\tau}}}\big[\text{KPI} (\mathbf{P}_0, \boldsymbol{\alpha}, \mathbf{H}_{\tau})\big]\bigg\}, \label{eq: optimization target}
\end{align}
where the objective function in \eqref{eq: optimization target} is expressed as the average KPI over the CSI distribution $\mathrm{P}_{\mathbf{H}_{\tau}}$ for configuration $\tau$. Examples of KPI include the sum-achievable rate, as it will be detailed in Sec. \ref{sec:numres}.

Intuitively, if $\alpha_c$ and $P_{0,c}$ are large, the intended received power at the BS of cell $c$ is high, but the interference generated to neighboring BSs is also significant. Conversely, if $\alpha_c$ and $P_{0,c}$ are small, both intended signal and interference are low. Therefore, the solution of problem \eqref{eq: optimization target} hinges on the identification of an optimized trade-off between intra-cell received power and inter-cell interference.

The objective in \eqref{eq: optimization target} cannot be directly evaluated, since it depends on the unknown distribution $\mathrm{P}_{\mathbf{H}_{\tau}}$. However, it can be estimated by using the CSI dataset $\mathcal{D}_{\tau}$ via the empirical average
\begin{align}
     f_{\tau}(\mathbf{P}_0,\boldsymbol{\alpha})= \frac{1}{S_{\tau}}\sum_{s=1}^{S_{\tau}} \text{KPI} (\mathbf{P}_0,\boldsymbol{\alpha}, \mathbf{H}_{\tau,s}),
\label{eq: noisy observation}
\end{align}
where we recall that $S_{\tau}$ is the number of available measurements $\{\mathbf{H}_{\tau,s}\}$ in dataset $\mathcal{D}_{\tau}$. Overall, the problem of interest is the optimization\begin{align}
    \max \limits_{\mathbf{P_0},\boldsymbol{\alpha}}  f_{\tau}(\mathbf{P}_0,\boldsymbol{\alpha}). \label{eq: poptimization target}
\end{align}
When one restricts the parameters $(\mathbf{P}_0,\boldsymbol{\alpha})$ as in Table \ref{table: olpc choices}, the problem is discrete.

One could solve the discrete optimization problem \eqref{eq: poptimization target} using exhaustive search, but this may not be computationally feasible. In fact, the optimization space includes $N_{OLPC}$ possible OLPC choices. In the next sections, we will explore more efficient, approximate solutions. As we will detail in Sec. \ref{sec: BMO}, the proposed meta-learning methods leverage the principle of transferring knowledge from previously encountered configurations $\tau$ in order to prepare to optimize power allocation for new configurations.

\begin{algorithm}[t!]
\caption{Bayesian Optimization (BO) for a given configuration $\tau$}\label{table: BO}
\SetKwInOut{Input}{Input}
\Input{GP prior $(\mu(\cdot),k(\cdot,\cdot))$, CSI dataset $\mathcal{D}_{\tau}$, maximum number of rounds $T_{max}$}
\SetKwInOut{Output}{Output}
\Output{Optimized $\mathbf{x}^*$}\
Initialize round $t=0$, empty matrix $\mathbf{X}_0=[\;]$, empty vector $\tilde{\mathbf{f}}_0=[\;]$\\
\While{\emph{not converged}}{
Obtain the next OLPC vector $\mathbf{x}_{t+1}$ using \eqref{eq: acquisition function}\\
Obtain observation $\tilde{f}_{t+1}\sim \mathcal{N}(\tilde{f}_{t+1}|f(\mathbf{x}_{t+1}),\sigma^2)$\\
Update matrix $\mathbf{X}_{t+1}=[\mathbf{X}_t,\mathbf{x}_{t+1}]$ and vector $\tilde{\mathbf{f}}_{t+1}=[\tilde{\mathbf{f}}_t,\tilde{f}_{t+1}]^{\sf T}$\\
Set $t=t+1$\\
}
Return $\mathbf{x}^*=\mathbf{x}_{t^*} \;\text{with} \; t^*=\arg \max_{t'\in \{1,...,t-1\}}\tilde{f}_{t'}$\\
\end{algorithm}

\section{Bayesian Optimization}
\label{sec: BO}
As a first approach to address the black-box optimization problem \eqref{eq: poptimization target}, we review the solution proposed in \cite{Maggi2021bogp}, which models the objective function $f_{\tau}(\mathbf{P}_0,\boldsymbol{\alpha})$ as a Gaussian Process (GP) and applies Bayesian optimization (BO). As we will detail, the approach is based on black-box evaluations of the KPI $f_{\tau}(\mathbf{P}_0,\boldsymbol{\alpha})$ at a sequence of trial solutions $(\mathbf{P}_0,\boldsymbol{\alpha})$. As anticipated, the approach does not require an explicit channel estimation step in order to address the optimization (\ref{eq: poptimization target}). To simplify the notation,  we remove the dependence on the configuration $\tau$, which is assumed to be fixed throughout this section. 

A GP is defined by a mean function $\mu(\cdot)$ and a kernel function $k(\cdot,\cdot)$. The kernel function may be chosen, for instance, as $k(\mathbf{x},\mathbf{x}')=\exp(-\gamma||\mathbf{x}-\mathbf{x}'||^2)$. Intuitively, the role of the kernel function is to quantify the similarity between input parameters $\mathbf{x}$ and $\mathbf{x}'$ in terms of the respective KPI values. Specifically, writing  $\mathbf{x}=[\mathbf{P}_0^{\sf T},\boldsymbol{\alpha}^{\sf T}]^{\sf T}$ for the vector of variables under optimization in problem \eqref{eq: poptimization target}, the GP prior on the objective function $f(\mathbf{P}_0,\boldsymbol{\alpha})=f(\mathbf{x})$ stipulates that, for any set of $T$ inputs $\mathbf{X}=[\mathbf{x}_1,\hdots,\mathbf{x}_T]$, the corresponding values $\mathbf{f}(\mathbf{X})=[f(\mathbf{x}_1),\hdots,f(\mathbf{x}_T)]^{\sf T}$ of the objective function are jointly distributed as
\begin{align}
    p(\mathbf{f}(\mathbf{X})=\mathbf{f})=\mathcal{N}(\mathbf{f}|\boldsymbol{\mu}(\mathbf{X}),\mathbf{K}(\mathbf{X})), \label{eq: GP prior}
\end{align}
where $\mathbf{f}=[f_1,...,f_T]^{\sf T}$ is a $T\times 1$ real vector; $\boldsymbol{\mu}(\mathbf{X})=[\mu(\mathbf{x}_1),...,\mu(\mathbf{x}_T)]^{\sf T}$ is the $T \times 1$ mean vector; and $\mathbf{K}(\mathbf{X})$ represents the $T\times T$ covariance matrix, whose $(t,t')$th entry is given as $[\mathbf{K}(\mathbf{X})]_{t,t'}=k(\mathbf{x}_t,\mathbf{x}_{t'})$ with $t,t'\in\{1,...,T\}$. By \eqref{eq: GP prior}, the mean function $\mu(\cdot)$ encodes prior knowledge about the values of the objective function for any fixed input $\mathbf{x}$, while the kernel encodes prior knowledge about the variability of the loss function across pairs of values of $\mathbf{x}$.

One may further assume that the KPI values obtained from the simulator or look-up table are noisy, yielding the vector of noisy observed KPI values $\tilde{\mathbf{f}}=[\tilde{f}_1,...,\tilde{f}_T]^{\sf T}$. Specifically, assuming the observation noise is Gaussian, we have the conditional distribution $p(\tilde{\mathbf{f}}|\mathbf{f})=\mathcal{N}(\tilde{\mathbf{f}}|\mathbf{f},\sigma^2\mathbf{I}_T)$, where $\sigma^2$ is the variance of the observation noise and $\mathbf{I}_T$ is the $T\times T$ identity matrix.

Using \eqref{eq: GP prior}, the posterior distribution of the objective value $f(\mathbf{x})$ at OLPC option $\mathbf{x}$ given the observation of previous input and KPI pairs $(\mathbf{X},\tilde{\mathbf{f}})$ can be obtained as \cite{Rasmussen2004}
\begin{equation}p(f(\mathbf{x})=f|\mathbf{X},\tilde{\mathbf{f}})= \mathcal{N}(f|\mu(\mathbf{x}|\mathbf{X},\tilde{\mathbf{f}}), \sigma^2(\mathbf{x}|\mathbf{X},\tilde{\mathbf{f}})),\label{eq: GP posterior}\end{equation} 
where
\begin{subequations}
    \begin{equation}
    \mu(\mathbf{x}|\mathbf{X},\tilde{\mathbf{f}})=\mu(\mathbf{x})+\tilde{\mathbf{k}}(\mathbf{x})^{\sf T}(\tilde{\mathbf{K}}(\mathbf{X}))^{-1}(\tilde{\mathbf{f}}-\boldsymbol{\mu}(\mathbf{X})),\label{eq: GP mean}\end{equation}
    \begin{equation} \sigma^2(\mathbf{x}|\mathbf{X},\tilde{\mathbf{f}})=k(\mathbf{x},\mathbf{x})-\tilde{\mathbf{k}}(\mathbf{x})^{\sf T}(\tilde{\mathbf{K}}(\mathbf{X}))^{-1}\tilde{\mathbf{k}}(\mathbf{x}),\label{eq: kernel}\end{equation}
\end{subequations}
with $\tilde{\mathbf{K}}(\mathbf{X})=\mathbf{K}(\mathbf{X})+\sigma^2\mathbf{I}_T$, and $\tilde{\mathbf{k}}(\mathbf{x})$ being the $T \times 1$ covariance vector $\tilde{\mathbf{k}}(\mathbf{x})=[k(\mathbf{x},\mathbf{x}_1),\hdots,k(\mathbf{x},\mathbf{x}_T)]^{\sf T}.$ The distribution \eqref{eq: GP posterior} can be used to obtain an estimated value of the objective $f(\mathbf{x})$ as the mean $\mu(\mathbf{x}|\mathbf{X},\tilde{\mathbf{f}})$, as well as to quantify the corresponding uncertainty of the estimate via the variance $\sigma^2(\mathbf{x}|\mathbf{X},\tilde{\mathbf{f}})$.

At each round $t$, Bayesian optimization leverages GP inference to optimize the selection of the next vector $\mathbf{x}_{t+1}$ of OLPC parameters at which to evaluate the KPI. This is done by maximizing an \textit{acquisition function} $F(\mathbf{x}_{t+1}|\mathbf{X}_t,\tilde{\mathbf{f}}_t)$, which depends on the previous observations $(\mathbf{X},\tilde{\mathbf{f}})=(\mathbf{x}_1,...,\mathbf{x}_t,\tilde{f}_1,...,\tilde{f}_t)$, via the optimization
\begin{align}
    \mathbf{x}_{t+1}=\arg \max \limits_{\mathbf{x}} F(\mathbf{x}|\mathbf{X}_t,\tilde{\mathbf{f}}_t),
\label{eq: acquisition function}
\end{align}
where the OLPC vector $\mathbf{x}$ is constrained to take the values in Table \ref{table: olpc choices}.

A standard example of acquisition function is the \textit{expected improvement function}, which computes the average positive increment in the function $f(\mathbf{x}_{t+1})$ evaluated at $\mathbf{x}_{t+1}$ based on \eqref{eq: GP posterior} \cite{huang2006sequential}. Defining as $f^*_t=\max\{\tilde{f}_1,...,\tilde{f}_t\}$ the current best observed objective value, the expected improvement function is defined as \cite{Maggi2021bogp}
\begin{align}
    F(\mathbf{x}|\mathbf{X},\tilde{\mathbf{f}})&=\Big[\mu(\mathbf{x}|\mathbf{X},\tilde{\mathbf{f}})-f^*_t-\xi\Big]\Phi(\delta)+\sigma^2(\mathbf{x}|\mathbf{X},\tilde{\mathbf{f}})\phi(\delta),\label{eq: EI func}
\end{align}
where
\begin{align}
    \delta &=\frac{\mu(\mathbf{x}|\mathbf{X},\tilde{\mathbf{f}})-f^*_t-\xi}{\sigma^2(\mathbf{x}|\mathbf{X},\tilde{\mathbf{f}})};
\end{align}
functions $\mu(\mathbf{x}|\mathbf{X},\tilde{\mathbf{f}})$ and $\sigma^2(\mathbf{x}|\mathbf{X},\tilde{\mathbf{f}})$ are given as in \eqref{eq: GP mean} and \eqref{eq: kernel}; $\xi\in[0,1)$ is an exploration parameter; and $\Phi(\cdot)$ and $\phi(\cdot)$ are the standard Gaussian cumulative and probability density function, respectively. For a risk-sensitive system with a well-specified GP prior, we may choose small $\xi$ (e.g., $\xi=0.01$ or even $\xi=0$). In contrast, where the prior is not tailored to the given problem, one can use larger values of $\xi$ to enable exploration \cite{Maggi2021bogp}.

The overall Bayesian optimization procedure is summarized in Algorithm \ref{table: BO}. As a convergence criterion, we can fix the number of iterations to some value $T_{max}$, or else stop when the expected improvement \eqref{eq: EI func} is small enough.

\begin{algorithm}[t!]
\caption{Bayesian Meta-Optimization (Meta-BO)}\label{table: BMO}
\SetKwInOut{Input}{Input}
\Input{Parameterized GP prior $(\mu_{\boldsymbol{\theta}}(\cdot),k_{\boldsymbol{\theta}}(\cdot,\cdot))$, meta-training data $\mathbf{X}_{1:N}$, $\tilde{\mathbf{f}}_{1:N}$, stepsize $\beta$}
\SetKwInOut{Output}{Output}
\Output{Optimized hyperparameters vector $\boldsymbol{\theta}^*$}\
Initialize hyperparameters vector $\boldsymbol{\theta}$\\
\While{\emph{not done}}{
Evaluate gradient $\nabla_{\boldsymbol{\theta}}\mathcal{L}(\boldsymbol{\theta}|\mathbf{X}_{1:N},\tilde{\mathbf{f}}_{1:N})$ using \eqref{eq: MLE derivatives}\\
Update hyper-parameters using gradient descent $\boldsymbol{\theta} \leftarrow \boldsymbol{\theta}-\beta \nabla_{\boldsymbol{\theta}}\mathcal{L}(\boldsymbol{\theta}|\mathbf{X}_{1:N},\tilde{\mathbf{f}}_{1:N})$\\
}
Return $\boldsymbol{\theta}^*$\\
Given a new network configuration $\tau$, apply BO with hyperparameter $\boldsymbol{\theta}^*$\\
\end{algorithm}

\section{Bayesian Meta-optimization}
\label{sec: BMO}
Solving problem \eqref{eq: poptimization target} separately for each configuration $\tau$ via Bayesian optimization (Algorithm \ref{table: BO}) may entail significant complexity in terms of number $S_{\tau}$ of required CSI samples, as well as number of evaluations of KPI values, i.e., the number of iterations in Algorithm \ref{table: BO}. In this section, we introduce Bayesian meta-optimization \cite{ivana2021meta,bing2021meta}, which uses offline data collected from multiple system configurations $\tau$ as a means to reduce optimization complexity when applied to any configuration $\tau$ at run time.

In Bayesian meta-optimization, we assume that, in an offline phase, we can collect data from $N$ configurations, denoted as $\tau_1,...,\tau_N$. These configurations may correspond to previous deployments or to concurrent deployments located elsewhere the system or to previous runs of a simulator with different settings, such as inter-site distances and number of UEs. For each configuration $\tau_n$, with $n=1,...,N$, we have access to a dataset $\mathcal{D}_{\tau_n}$ of $S_{\tau_n}$ CSI samples, which can be used to obtain the objective function $f_{\tau_n}(\mathbf{x})$ in \eqref{eq: noisy observation}. Furthermore, for each task $\tau_n$, we assume to have collected $T_n$ inputs $\mathbf{X}_n=[\mathbf{x}_{n,1},...,\mathbf{x}_{n,T_n}]$, as well as the corresponding noisy observations $\tilde{\mathbf{f}}_n=[\tilde{f}_{n,1},...,\tilde{f}_{n,T_n}]$ of the actual objective values $f_{n,t}=f_{\tau_n}(\mathbf{x}_{n,t})$. We refer to the above collected data available from $N$ configurations as \textit{meta-training data}. In practice, the designer may equivalently only have access to $T_n$ evaluations of the KPI function. In our experiments, we explore values $T_n$ in the range [1,30]. We aim at using these data to improve efficiency on new tasks sampled from the same environment.

To this end, Bayesian meta-optimization uses meta-training data to optimize the GP prior via parametric mean function $\mu_{\boldsymbol{\theta}}(\cdot)$ and kernel function $k_{\boldsymbol{\theta}}(\cdot)$, which are functions of a vector of \textit{hyperparameters} $\boldsymbol{\theta}$. Specifically, we consider the parametric kernel function \cite{rothfuss2021pacoh}
\begin{equation}
k_{\boldsymbol{\theta}}(\mathbf{x},\mathbf{x}')=\exp{(-||\psi_{\boldsymbol{\theta}}(\mathbf{x})-\psi_{\boldsymbol{\theta}}(\mathbf{x}')||_2^2}),\label{eq: parametric function}
\end{equation}
where $\psi_{\boldsymbol{\theta}}(\cdot)$ is a neural network with hyperparameter vector $\boldsymbol{\theta}\in\mathbb{R}^L$ constituting its synaptic weights and biases and we also assume $\mu_{\boldsymbol{\theta}}(\mathbf{x})$ to be a neural network. By optimizing the GP prior via \eqref{eq: parametric function}, the goal is to ensure that Bayesian optimization applied to a new configuration $\tau$ can produce an effective solution with fewer samples $S_{\tau}$ and fewer evaluations $T_{max}$ of the KPI.

Intuitively, the role of the kernel function is to quantify the similarity between power control parameters $\mathbf{x}$ and $\mathbf{x}'$ in terms of the respective KPI values obtained for a given configuration. The standard approach in BO is to select this kernel as a predefined distance metric, e.g., the Euclidean distance in \cite{Maggi2021bogp}, which may not reflect well the specific properties of the given optimization problem \eqref{eq: poptimization target}. In contrast, Bayesian meta-optimization aims at optimizing the kernel function so as to account for the structure of the power control optimization problems \eqref{eq: poptimization target} for the $N$ configurations for which we have meta-training data. The rationale is that one expects such structure to be sufficiently related to that of any new configuration $\tau$ of interest.

Bayesian meta-optimization, is formulated by introducing \textit{meta-training loss} incurred on the meta-training data $\mathbf{X}_{1:N}=[\mathbf{X}_1,...,\mathbf{X}_{N}]$ and $\tilde{\mathbf{f}}_{1:N}=[\tilde{\mathbf{f}}_1,...,\tilde{\mathbf{f}}_N]$ when using hyperparameter vector $\boldsymbol{\theta}$ as
\begin{align}
    \mathcal{L}(\boldsymbol{\theta}|\mathbf{X}_{1:N},\tilde{\mathbf{f}}_{1:N})=- \frac{1}{N} \sum_{n=1}^N \frac{1}{T_n}\ln p_{\boldsymbol{\theta}}(\tilde{\mathbf{f}}_{n}|\mathbf{X}_n), \label{eq: meta-traning loss}
\end{align}
where
\begin{align}
    &\ln p_{\boldsymbol{\theta}}(\tilde{\mathbf{f}}_{n}|\mathbf{X}_n)\nonumber \\ &=-\frac{1}{2}\Big(\tilde{\mathbf{f}}_n-\boldsymbol{\mu_{\theta}}(\mathbf{X}_n)\Big)^{\sf T}\Big(\tilde{\mathbf{K}}_{\boldsymbol{\theta}}(\mathbf{X}_n)\Big)^{-1}\Big(\tilde{\mathbf{f}}_n-\boldsymbol{\mu_{\theta}}(\mathbf{X}_n)\Big)\nonumber \\ &-\frac{1}{2}\ln{\Big|\tilde{\mathbf{K}}_{\boldsymbol{\theta}}(\mathbf{X}_n)\Big|}-\frac{T_n}{2}\ln{(2\pi)},\label{eq: meta likelihood}
\end{align}
with $\boldsymbol{\mu_{\theta}}(\mathbf{X}_n)=[\mu_{\boldsymbol{\theta}}(\mathbf{x}_{n,1}),...,\mu_{\boldsymbol{\theta}}(\mathbf{x}_{n,T_n})]^{\sf T}$; $[\mathbf{K}_{\boldsymbol{\theta}}(\mathbf{X}_n)]_{t,t'}=k_{\boldsymbol{\theta}}(\mathbf{x}_{n,t},\mathbf{x}_{n,t'})$ for $(t,t')\in\{1,...,T_n\}$; and $\tilde{\mathbf{K}}_{\boldsymbol{\theta}}(\mathbf{X}_n)=\mathbf{K}_{\boldsymbol{\theta}}(\mathbf{X}_n)+\sigma^2\mathbf{I}_{T_n}$. The meta-training loss \eqref{eq: meta-traning loss} is the empirical average of the negative log-likelihood evaluated on the meta-training data \cite{rothfuss2021pacoh}. The optimal hyperparameter $\boldsymbol{\theta}^*$ is obtained by addressing the problem
\begin{align}
    \boldsymbol{\theta}^*=\arg \min \limits_{\boldsymbol{\theta}} \mathcal{L}(\boldsymbol{\theta}|\mathbf{X}_{1:N},\tilde{\mathbf{f}}_{1:N}). \label{eq: meta mle}
\end{align}

To implement the optimization in \eqref{eq: meta mle}, we adopt a gradient-based optimizer. The partial derivative of the meta-training loss with respect to the $j$-th component $\theta_j$ of the hyperparameters vector $\boldsymbol{\theta}$ is computed as
\begin{align}
    &\frac{\partial}{\partial \theta_j} \mathcal{L}(\boldsymbol{\theta}|\mathbf{X}_{1:N},\tilde{\mathbf{f}}_{1:N})\nonumber \\ &=-\frac{1}{N}\sum_{n=1}^N\Bigg(\frac{1}{2}\Big(\tilde{\mathbf{f}}_n-\boldsymbol{\mu_{\theta}}(\mathbf{X}_n)\Big)^{\sf T}\Big(\tilde{\mathbf{K}}_{\boldsymbol{\theta}}(\mathbf{X}_n)\Big)^{-1}\nonumber \\&\frac{\partial \tilde{\mathbf{K}}_{\boldsymbol{\theta}}(\mathbf{X}_n)}{\partial \theta_j}\Big(\tilde{\mathbf{K}}_{\boldsymbol{\theta}}(\mathbf{X}_n)\Big)^{-1}\Big(\tilde{\mathbf{f}}_n-\boldsymbol{\mu_{\theta}}(\mathbf{X}_n)\Big) \nonumber \\ &-\frac{1}{2}\text{tr}\bigg(\tilde{\mathbf{K}}_{\boldsymbol{\theta}}(\mathbf{X}_n)^{-1}\frac{\partial \tilde{\mathbf{K}}_{\boldsymbol{\theta}}(\mathbf{X}_n)}{\partial \theta_j}\bigg)\Bigg)\frac{1}{T_n} \nonumber \\ 
    &=-\frac{1}{N}\sum_{n=1}^N\frac{1}{2T_n}\text{tr}\bigg(\Big(\boldsymbol{\Lambda\Lambda}^{\sf T}-\tilde{\mathbf{K}}_{\boldsymbol{\theta}}(\mathbf{X}_n)^{-1}\Big)\frac{\partial \tilde{\mathbf{K}}_{\boldsymbol{\theta}}(\mathbf{X}_n)}{\partial \theta_j}\bigg), \label{eq: MLE derivatives}
\end{align}
where $\boldsymbol{\Lambda}=\tilde{\mathbf{K}}_{\boldsymbol{\theta}}(\mathbf{X}_n)^{-1}(\tilde{\mathbf{f}}_n-\boldsymbol{\mu_{\theta}}(\mathbf{X}_n))$. The partial derivative term in \eqref{eq: MLE derivatives} can be estimated by backprop with the parameters in \eqref{eq: parametric function}.

The hyper-parameter $\boldsymbol{\theta}^*$ optimized with the gradient-based procedure outlined above is used to define the GP prior to be used for Bayesian optimization in new configurations for the purpose of improving the efficiency of Bayesian optimization. Overall, Bayesian meta-optimization is summarized in Algorithm \ref{table: BMO}.

\section{Bandit Optimization and Meta-optimization}\label{sec: mab}
Given the discrete nature of problem \eqref{eq: poptimization target} when considering Table \ref{table: olpc choices}, it can be directly modelled as a stochastic multi-armed bandit (MAB) model rather than as GP, which assumes continuous variables. In the MAB formulation, the total number of arms equals the number, $N_{OLPC}$, of OLPC parameters options listed in Table \ref{table: olpc choices}. The goal is to design a policy that selects the best ``arm'' i.e., the OLPC pair $(\mathbf{P}_0,\boldsymbol{\alpha})$ that optimizes problem \eqref{eq: poptimization target} after a small number of attempts. In practice, as in the case of Bayesian optimization, one accepts sub-optimal solutions that performs well enough.

\subsection{Bandit Policy}\label{ssec: bandit policy}
As in Bayesian optimization (see Algorithm \ref{table: BO}), for a configuration $\tau$, at the $t$th optimization round, the learning agent selects an OLPC configuration $\mathbf{x}_{t}$ from Table \ref{table: olpc choices} and observes a noisy version $\tilde{f}_{t}$ of the corresponding KPI value $f_{\tau}(\mathbf{x}_t)$. In a manner similar to \eqref{eq: acquisition function}, a bandit optimization policy maps the history $(\mathbf{X}_t,\tilde{\mathbf{f}}_t)$ of previous selections and corresponding cost functions up to round $t$ to the next selection $\mathbf{x}_{t+1}$. Specifically, we consider a stochastic bandit policy $p_{\omega}(\mathbf{x}|\mathbf{X}_t,\tilde{\mathbf{f}}_t)$, parameterized by a scalar $\omega\in [0,1]$, that defines the probability of selecting an OLPC configuration $\mathbf{x}$ at $t$-th round given the past history $(\mathbf{X}_t,\tilde{\mathbf{f}}_t)$. 
Policy $p_{\omega}(\mathbf{x}|\mathbf{X}_t,\tilde{\mathbf{f}}_t)$ can be defined via a recurrent neural network \cite{Craig2020nips} or via simpler functions such as the Exp3 policy in \cite{mackay2003}.

In this work, we consider the following \textit{modified Exp3 policy} 
\begin{align}
    p_{\omega}(\mathbf{x}|\mathbf{X}_t,\tilde{\mathbf{f}}_t)=(1-\omega)\frac{\exp(G(\mathbf{x},t-1))}{\sum_{\mathbf{x}'}G(\mathbf{x}',t-1)}+\frac{\omega}{N_{OLPC}},\label{eq: exp3}
\end{align}
where $\omega \in [0,1]$ is the policy parameter; the sum is over all the possible OLPC configurations in Table \ref{table: olpc choices}; and
\begin{align}
    G(\mathbf{x},t-1)=\sum_{i=1}^{t-1}k(\mathbf{x}_i,\mathbf{x})[p_{\omega}(\mathbf{x}|\mathbf{X}_{t-1},\tilde{\mathbf{f}}_{t-1})]^{-1}\tilde{f}_i, \label{eq: mab weight}
\end{align}
is a weighted average of the noisy objective function values obtained for input  $\mathbf{x}$ in the previous $t-1$ rounds, with $k(\cdot,\cdot)$ being a kernel function. While the conventional choice for the kernel function is the identity function $k(\mathbf{x},\mathbf{x}')=1$ if $\mathbf{x}=\mathbf{x}'$ and $k(\mathbf{x},\mathbf{x}')=0$ otherwise, here we will allow for a more general solution. This will be useful in the next subsection to facilitate the application of meta-learning.

Standard bandit optimization considers a fixed parameter parameter $\omega$, and is summarized in Algorithm \ref{table: MAB}.
\begin{algorithm}[t]
\caption{Multi-Armed Bandit Optimization (MAB) for a given configuration $\tau$}\label{table: MAB}
\SetKwInOut{Input}{Input}
\Input{Policy parameter $\omega$, CSI dataset $\mathcal{D}_{\tau}$, maximum number of rounds $T_{max}$}
\SetKwInOut{Output}{Output}
\Output{Optimized $\mathbf{x}^*$}\
Initialize round $t=0$, empty matrix $\mathbf{X}_0=[\;]$, empty vector $\tilde{\mathbf{f}}_0=[\;]$\\
\While{\emph{not converged}}{
Sample from policy $p_{\omega}(\mathbf{x}|\mathbf{X}_t,\tilde{\mathbf{f}}_t)$ to obtain $\mathbf{x}_{t+1}$\\
Obtain observation $\tilde{f}_{t+1}\sim \mathcal{N}(\tilde{f}_{t+1}|f(\mathbf{x}_{t+1}),\sigma^2)$\\
Update matrix $\mathbf{X}_{t+1}=[\mathbf{X}_t,\mathbf{x}_{t+1}]$ and vector $\tilde{\mathbf{f}}_{t+1}=[\tilde{\mathbf{f}}_t,\tilde{f}_{t+1}]^{\sf T}$\\
Set $t=t+1$\\
}
Return $\mathbf{x}^*=\arg \max_{t'\in\{ 1,...,t-1\}}\tilde{f}_{t'}$\\
\end{algorithm}

\subsection{Bandit Meta-Optimization}\label{ssec: meta mab}
Following the meta-learning setting introduced in Sec. \ref{sec: BMO}, in this section we propose a bandit meta-optimization strategy. As in Sec. \ref{sec: BMO}, we assume availability of data for $N$ system configurations. The goal of bandit meta-optimization is to use such meta-training data, given by $\mathbf{X}_{1:N}$ and $\tilde{\mathbf{f}}_{1:N}$ as defined in the previous sections, to optimize a hyperparameter vector defining the bandit policy.

To this end, we propose to instantiate the kernel function $k_{\boldsymbol{\varphi}}(\cdot,\cdot)$ in the Exp3 policy \eqref{eq: exp3} as in \eqref{eq: parametric function} with neural network parameters $\boldsymbol{\varphi}$. We aim at optimizing the parameter tuple $\boldsymbol{\theta}=(\boldsymbol{\varphi},\omega)$ defining the resulting policy $p_{\boldsymbol{\theta}}(\mathbf{x}|\mathbf{X}_n,\tilde{\mathbf{f}}_n)$ to ensure that bandit meta-optimization applied to a new configuration $\tau$ can select an effective OLPC vector with a smaller number of trials. 

To this end, we define the following meta-training loss as
\begin{align}
    \mathcal{L}(\boldsymbol{\theta}|\mathbf{X}_{1:N},\tilde{\mathbf{f}}_{1:N})=-\frac{1}{N}\sum_{n=1}^N\mathbb{E}_{\mathbf{x} \sim p_{\boldsymbol{\theta}}(\cdot|\mathbf{X}_n,\tilde{\mathbf{f}}_n)}[\tilde{f}(\mathbf{x})],\label{eq: meta mab loss}
\end{align}
where the expectation is taken with respect to the bandit policy $p_{\boldsymbol{\theta}}(\mathbf{x}|\mathbf{X}_n,\tilde{\mathbf{f}}_n)$ based on the available history $(\mathbf{X}_n,\tilde{\mathbf{f}}_n)$ of observations for each $n$-th configuration $\tau_n$. To implement the optimization over \eqref{eq: meta mab loss}, we adopt a gradient-based optimizer. The gradient of the meta-training loss with respect to policy vector $\boldsymbol{\varphi}$ and $\omega$ is evaluated as \cite{Craig2020nips}
\begin{align}
    &\nabla_{\boldsymbol{\theta}}\mathcal{L}(\boldsymbol{\theta}|\mathbf{X}_{1:N},\tilde{\mathbf{f}}_{1:N})\nonumber \\&=-\frac{1}{N}\sum_{n=1}^N\mathbb{E}_{\mathbf{x} \sim p_{\boldsymbol{\theta}}(\cdot|\mathbf{X}_n,\tilde{\mathbf{f}}_n)}\bigg[\tilde{f}(\mathbf{x})\nabla_{\boldsymbol{\theta}}\log p_{\boldsymbol{\theta}}(\mathbf{x}|\mathbf{X}_n,\tilde{\mathbf{f}}_n)\bigg].\label{eq: meta mab gradient}
\end{align}

The meta-learned optimal policy vector $\boldsymbol{\theta}^*=(\boldsymbol{\varphi}^*,\omega^*)$ is then used in the bandit policy used in Algorithm \ref{table: MAB} to optimize the OLPC variables for a new configuration. Bandit meta-optimization is summarized in Algorithm \ref{table: meta mab}.
\begin{algorithm}[t!]
\caption{Bandit Meta-Optimization (Meta-MAB)}\label{table: meta mab}
\SetKwInOut{Input}{Input}
\Input{Parameterized policy $p_{\boldsymbol{\theta}}(\mathbf{x}|\mathbf{X}_n,\tilde{\mathbf{f}}_n)$, meta-training data $\mathbf{X}_{1:N}$, $\tilde{\mathbf{f}}_{1:N}$, stepsize $\eta$}
\SetKwInOut{Output}{Output}
\Output{Optimized policy vector $\boldsymbol{\theta}^*=(\boldsymbol{\varphi}^*$, $\omega^*)$}\
Initialize policy vector $\boldsymbol{\theta}=(\boldsymbol{\varphi},\omega)$\\
\While{\emph{not done}}{

Evaluate gradient $\nabla_{\boldsymbol{\theta}}\mathcal{L}(\boldsymbol{\theta}|\mathbf{X}_{1:N},\tilde{\mathbf{f}}_{1:N})$ using \eqref{eq: meta mab gradient}\\
Update policy vector using gradient descent $\boldsymbol{\theta} \leftarrow \boldsymbol{\theta}-\eta \nabla_{\boldsymbol{\theta}}\mathcal{L}(\boldsymbol{\theta}|\mathbf{X}_{1:N},\tilde{\mathbf{f}}_{1:N})$\;}
Return $\boldsymbol{\theta}^*$\\
Given a new network configuration $\tau$, apply MAB (Algorithm \ref{table: MAB}) with hyperparameter $\boldsymbol{\theta}^*=(\boldsymbol{\varphi}^*,\omega^*)$ with kernel function $k_{\boldsymbol{\varphi}}(\cdot,\cdot)$\\
\end{algorithm}

\section{Contextual Bayesian and Bandit Meta-Optimization}\label{sec: context}
In the previous sections, we have assumed that no information is available about the current configuration $\tau$ apart from the CSI dataset $\mathcal{D}_{\tau}$. In practice, the system may have access to \textit{context} information about the deployment underlying the configuration $\tau$, such as the geometric layout, expected UE positions, or the fading statistics. In this section, we introduce a generalization of the meta-optimization strategies described in Sec. \ref{sec: BMO} and Sec. \ref{sec: mab} that can leverage configuration-specific context information to optimize OLPC parameters $\mathbf{x}=(\mathbf{P}_0,\boldsymbol{\alpha})$.

\subsection{Context-Based Meta-Optimization}\label{ssec: context bmo}
Let $\mathbf{c}_{\tau}$ denote a context vector specific to configuration $\tau$, which includes all the information available at the optimizer about configuration $\tau$. The key idea of the proposed methods is to use meta-training data from multiple tasks in order to optimize a procedure that can adapt the parameters $\boldsymbol{\theta}$ for BO or MAB optimization to the configuration-specific context $\mathbf{c}_{\tau}$.

Formally, for each meta-training configuration $\tau_n$, we have access to data $(\mathbf{X}_n,\tilde{\mathbf{f}}_n,\mathbf{c}_n)$, where $\mathbf{c}_n$ is the context vector for the meta-training task $\tau_n$. Therefore, as compared to the meta-learning settings studied in the last two sections, here we assume the additional availability of the context vector $\mathbf{c}_n$ for each task $\tau_n$. Accordingly,  at run time, the optimizer is given context vector $\mathbf{c}_{\tau}$ for the current configuration $\tau$. The goal is to effectively adapt the optimizer's parameters $\boldsymbol{\theta}$ to the context vector $\mathbf{c}_{\tau}$ by leveraging knowledge transferred from the meta-learning tasks.

The proposed approach leverages meta-learning data to optimize a parametric mapping $q_{\mathbf{V}}(\cdot)$ between context $\mathbf{c}_{\tau}$ and parameters $\boldsymbol{\theta}$. The mapping depends on a parameter matrix $\mathbf{V}$ that is to be optimized based on meta-training data. Once vector $\mathbf{V}$, and hence also the parametric mapping $q_{\mathbf{V}}(\cdot)$, are fixed, an optimized per-task configuration hyperparameters $\boldsymbol{\theta}^*_{\tau}$ is obtained as $\boldsymbol{\theta}^*_{\tau}=q_{\mathbf{V}}(\mathbf{c}_{\tau})$ for the new task $\tau$.

Intuitively, an effective mapping $q_{\mathbf{V}}(\cdot)$ should map similar context vectors, defining similar configurations, into similar parameter vectors. Two context vectors are similar if the respective KPIs depend in an analogous way on the parameters $(\mathbf{P}_0,\boldsymbol{\alpha})$ under optimizations. Since, as we will detail in the next subsection, the context vector typically encodes information about the topology of the network, the mapping should account for the extent to which  topologies with similar characteristics call for related optimized power control parameters $\mathbf{x}$. 

In order to facilitate the optimization of mapping functions with this intuitive property, we propose here to adopt the linear function\begin{align}
    q_{\mathbf{V}}(\mathbf{c})=\sum_{n=1}^N \kappa(\mathbf{c},\mathbf{c}_n)\boldsymbol{\nu}_n,\label{eq:linear regressor}
\end{align} where we have introduced the \emph{context kernel} function $\kappa(\mathbf{c},\mathbf{c}')$ to measure the similarity between two context vectors $\mathbf{c}$ and $\mathbf{c}'$. As detailed in the next subsection, the context kernel function is set by the optimizer to capture the desired similarity properties between two context vectors. The mapping (\ref{eq:linear regressor}) depends on parameter vectors $\boldsymbol{\nu}_1,...,\boldsymbol{\nu}_N$ of the same dimension of the parameter vector $\boldsymbol{\theta}$, which we collect in the parameter matrix  $\mathbf{V}=[\boldsymbol{\nu}_1,...,\boldsymbol{\nu}_N]$ to be optimized. Finally, introducing the vector  $\boldsymbol{\kappa}(\mathbf{c})=[\kappa(\mathbf{c},\mathbf{c}_1),...,\kappa(\mathbf{c},\mathbf{c}_N)]^{\sf T}$, the mapping (\ref{eq:linear regressor}) can be expressed as\begin{align}
    q_{\mathbf{V}}(\mathbf{c})=\mathbf{V}\boldsymbol{\kappa}(\mathbf{c}).\label{eq:linear regressor1}
\end{align}

By \eqref{eq:linear regressor}, or \eqref{eq:linear regressor1}, the parameter vector \begin{equation}\label{eq:contpar}\boldsymbol{\theta}^*_{\tau}=q_{\mathbf{V}}(\mathbf{c}_{\tau})\end{equation} for the test configuration $\tau$ is modelled as a linear combination of vectors $\boldsymbol{\nu}_n$, with each vector $\boldsymbol{\nu}_n$ being weighted by the similarity $\kappa(\mathbf{c}_{\tau},\mathbf{c}_n)$ between context vectors $\mathbf{c}_{\tau}$ and $\mathbf{c}_n$. Implementing the intuition detailed at the beginning of this subsection, we can view $\boldsymbol{\nu}_n$ as the parameter vector assigned to the meta-learning configuration $\tau_n$, and the parameter vector $\boldsymbol{\theta}^*_{\tau}$  in (\ref{eq:contpar}) as being closer to vectors $\boldsymbol{\nu}_n$ corresponding to more similar configurations $\tau_n$ according to the kernel similarity measure $\kappa(\mathbf{c}_\tau,\mathbf{c}_n)$.

The parameter matrix $\mathbf{V}$ has a number of entries equal to the product of the size of model parameter $\theta$, denoted as $L$,  and the number $N$ of meta-learning tasks. This may be exceedingly large, causing optimization during meta-learning to possibly overfit the meta-training data yielding poor performance on the test configuration. To address this problem, we propose to factorize the $L\times N$ matrix $\mathbf{V}$ by using a low-rank decomposition into two lower-dimensionality factors. Accordingly, we write the mapping  (\ref{eq:contpar}) as \begin{equation}\label{eq:redpar}\boldsymbol{\theta}^*_{\tau}=q_{\mathbf{V}_1,\mathbf{V}_2}(\mathbf{c}_{\tau})=\mathbf{V}_1\mathbf{V}_2^{\sf T}\boldsymbol{\kappa}(\mathbf{c}),\end{equation} which depends on the parameter matrices $\mathbf{V}_1\in\mathbb{R}^{L\times r}$ and $\mathbf{V}_2 \in\mathbb{R}^{N\times r}$ for rank $r<\min\{L,N\}$ being a hyperparameter.

\subsection{Context Graph Kernel}
\label{ssec: context kernel}
The choice of the \textit{context kernel} $\kappa(\cdot,\cdot)$ depends on the type of information included in the context vector for  each configuration. In this subsection, we introduce a solution that applies to the common situation in which the context vector includes information about the topology of the network, namely all distances between BSs and UEs. This setting is selected to demonstrate the importance of leveraging the structure inherent in the context vector, along with the corresponding symmetry properties of the mapping from context vector to model parameters. This is detailed next.

For the purpose of  power allocation, information about the topology of the network is important insofar as it determines the interference pattern among the links. In particular, the order in which the links are listed in the context vector $\mathbf{c}_{\tau}$ is not relevant. This implies that the mapping (\ref{eq:redpar}) should be invariant to permutations of the entries of the context vector. To enforce this invariance property,  we adopt the framework of \emph{graph kernels}  \cite{deepgraphkernel}.

To this end, we summarize information about topology of the network for configuration $\tau$ by means of an annotated interference graph $\mathcal{G}_{\tau}$ that retains information about within-cell UE-BS distances (see, e.g., \cite{he2021overview}). As illustrated in  Fig. \ref{fig: context graph}, in the \emph{interference graph} $\mathcal{G}_{\tau}$, each node represents a link between a UE and the serving BS. Each node $i$ is annotated with distance $d_i$ between the corresponding UE, also indexed by $i$ as UE-$i$, and the serving BS. A directed edge from node $i$ to node $j$ is included in graph $\mathcal{G}_{\tau}$ if the interference from the link associated with node $i$ to the link associated with node $j$ is sufficiently large. To gauge the level of interference from link $i$ to link $j$, we consider the distance $d_{ij}$ between UE-$i$ and the BS serving UE-$j$. If the ratio $d_{ij}/d_j$ of this distance to the distance between UE-$j$ and the serving BS is below some threshold, a directed edge is added between node $i$ and $j$.

The context kernel $\kappa(\mathbf{c}_\tau,\mathbf{c}_{\tau^{\prime}})$ is designed to measure the similarity between the graphs $\mathcal{G}_{\tau}$ and $\mathcal{G}_{\tau^{\prime}}$ corresponding to context vectors $\mathbf{c}_{\tau}$ and $\mathbf{c}_{\tau^{\prime}}$, respectively. There are a number of graph kernels that one can choose from for this purpose, ranging from graphlet kernels to deep graph kernels \cite{deepgraphkernel}. In this work, we focus on \emph{graphlet kernels} \cite{graphlet}, which are defined as
\begin{align}
    \kappa(\mathbf{c}_\tau,\mathbf{c}_{\tau^{\prime}})=\frac{\Psi(\mathcal{G}_\tau)^{\sf T}\Psi(\mathcal{G}_{\tau^{\prime}})}{||\Psi(\mathcal{G}_\tau)||_2||\Psi(\mathcal{G}_{\tau^{\prime}})||_2},\label{eq: graphlet kernel}
\end{align}
where $\Psi(\mathcal{G})$ is a vector of features extracted from the graph $\mathcal{G}$. Each such feature of vector $\Psi(\mathcal{G})$ counts the number of times a certain sub-graph is contained in the graph $\mathcal{G}$. We specifically propose to consider the following feature vector
\begin{align}
    &\Psi(\mathcal{G})=[\Psi_1(\mathcal{G}),...,\Psi_{N_U-1}(\mathcal{G})]^{\sf T},
     \label{eq:feature vector}
\end{align} where $\Psi_i(\mathcal{G})=$ number of nodes with in-degree equal to $i$. The rationale for this choice is that interference graphs with similar connectivity, as quantified by vector \eqref{eq:feature vector}, should also have similar characteristics in terms of the impact of power control decisions on interference levels. Accordingly, context vectors with a large value of the kernel \eqref{eq: graphlet kernel} are expected to have similar optimized power control parameters. Note that vector $\Psi(\mathcal{G})$ contains a number of entries equal to the number $N_U$ of UEs minus $1$, which corresponds to the number of nodes in the interference graph $\mathcal{G}$. Furthermore, the in-degree of a node is the number of incoming edges.

\subsection{Context-Based Bayesian Meta-Optimization}
\label{ssec: cbmo}
To define context-based Bayesian meta-optimization, we directly modify the meta-training loss introduced in Sec. \ref{sec: BMO} in \eqref{eq: meta-traning loss} for Bayesian meta-optimization as \begin{align}
    &\mathcal{L}(\mathbf{V}_1,\mathbf{V}_2|\mathbf{X}_{1:N},\tilde{\mathbf{f}}_{1:N},\mathbf{c}_{1:N})\nonumber \\&=-\frac{1}{N}\sum_{n=1}^N \frac{1}{T_n}\ln p_{q_{\mathbf{V}_1\mathbf{V}_2}(\mathbf{c}_n)}(\tilde{\mathbf{f}}_n|\mathbf{X}_n). \label{eq: context meta loss}
\end{align} The key difference is that the meta-training loss is now a function of the two matrix factors $\mathbf{V}_1$ and $\mathbf{V}_2$, rather  than being a function directly of the parameter vector $\boldsymbol{\theta}$. In fact, the parameter $\boldsymbol{\theta}$ is adapted to the context $\mathbf{c}_n$ of each task $\tau_n$ via the mapping $q_{\mathbf{V}_1\mathbf{V}_2}(\mathbf{c}_n)$.
The meta-learned optimal parameter matrices $\mathbf{V}_1^*$ and $\mathbf{V}^*_2$ are  obtained as the minimizer
\begin{align}
    (\mathbf{V}_1^*,\mathbf{V}_2^{\sf *T})=\arg \min \limits_{\mathbf{V}_1,\mathbf{V}_2}\mathcal{L}(\mathbf{V}_1,\mathbf{V}_2|\mathbf{X}_{1:N},\tilde{\mathbf{f}}_{1:N},\mathbf{c}_{1:N}),\label{eq: optimal Theta}
\end{align}where the optimization can be addressed via gradient-descent and backprop in a manner similar to problem \eqref{eq: meta mle}. 

\begin{figure}[t]

  \centering
  \centerline{\includegraphics[scale=0.30]{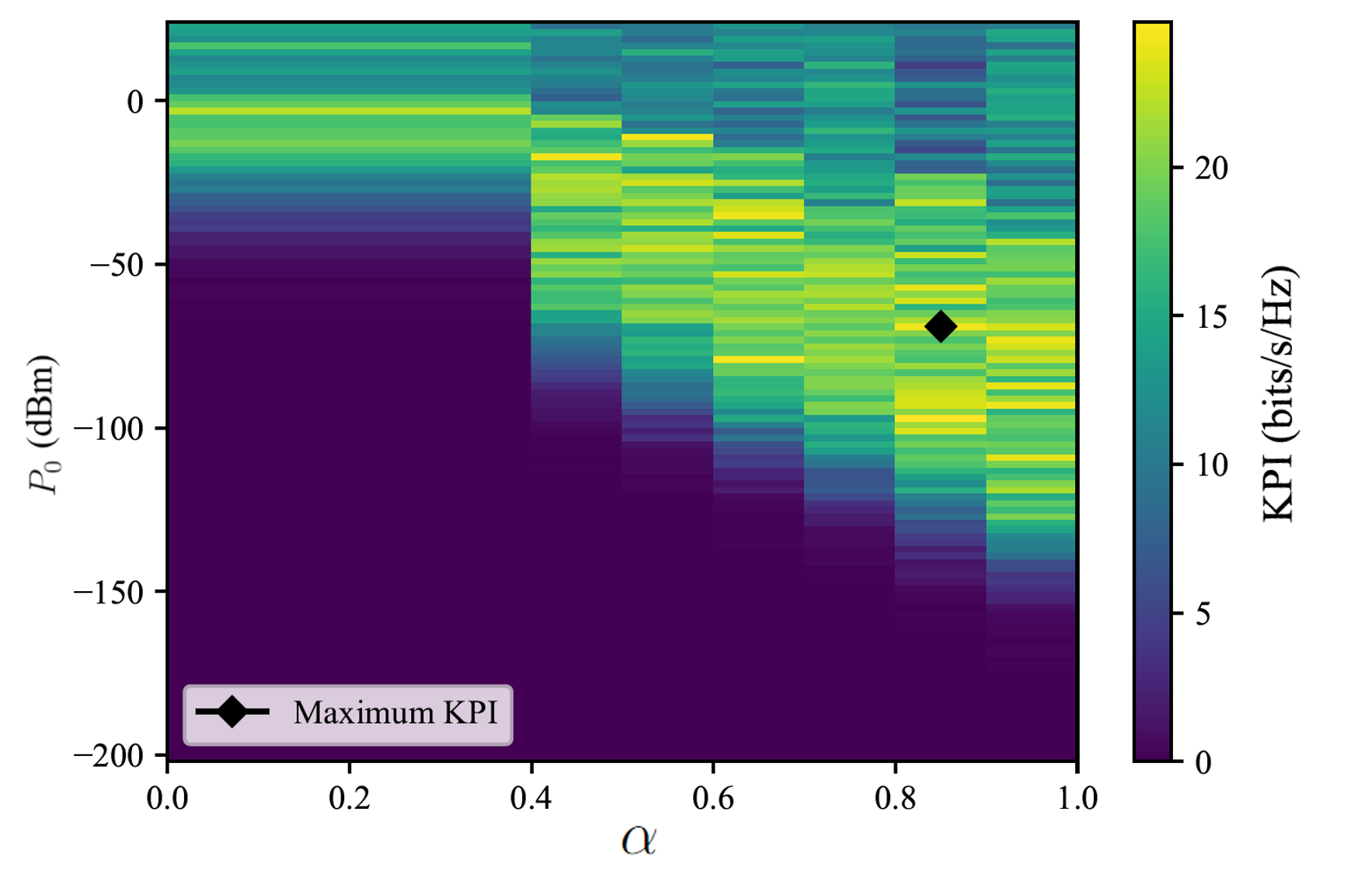}}
  \caption{Illustration of the objective function \eqref{eq: mimo kpi} for a given configuration $\tau$ in the optimization space $(P_0,\alpha)$ for the multi-cell system considered in Sec. \ref{sec:numres}.}
  \label{fig: exhaustive search}
\vspace{0cm}
\end{figure}

\subsection{Context-Based Bandit Meta-Optimization}
\label{ssec: cmmab}
In a similar way, context-based bandit meta-learning addresses the minimization of the meta-training loss obtained by  replacing in  \eqref{eq: meta mab loss}  the model parameter vector $\boldsymbol{\theta}$ with the output of $q_{\mathbf{V}_1\mathbf{V}_2}(\mathbf{c}_n)$ of the meta-trained mapping for each task $\tau_n$.  This yields the objective \begin{align}
    &\mathcal{L}(\mathbf{V}_1,\mathbf{V}_2|\mathbf{X}_{1:N},\tilde{\mathbf{f}}_{1:N},\mathbf{c}_{1:N})\nonumber \\&=\frac{1}{N}\sum_{n=1}^N\mathbb{E}_{\mathbf{x} \sim p_{q_{\mathbf{V}_1\mathbf{V}_2}(\mathbf{c}_n)}(\cdot|\mathbf{X}_n,\tilde{\mathbf{f}}_n)}[\tilde{f}(\mathbf{x})],\label{eq: context meta mab loss}
\end{align}
which can be addressed via gradient descent.

\section{Numerical Results}\label{sec:numres}
In this section, we present a number of experimental results with the goal of validating the potential benefits of the proposed meta-learning and contextual meta-learning methods for uplink power allocation via Bayesian  and bandit optimization.

\subsection{Setting}\label{ssec: params setting}
We consider a multi-cell MIMO system with a wrap-around radio distance model, in which  $N_U$ UEs in each cell are equipped with $N_T$ transmit antennas each, while the BSs serving the UEs in each cell are equipped with $N_R$ receiving antennas. Focusing on a single resource block, the CSI $\mathbf{H}_{\tau}$ consists of the $N_R\times N_T$ channel matrices $\mathbf{H}_{\tau,c,u,c'}$ describing the propagation channel between the $N_T$ antennas of the $u$th UE in cell $c$ and the $N_R$ antennas of the BS in cell $c'$. The KPI function in \eqref{eq: poptimization target} is instantiated as the sum of the spectral efficiencies for all users in the system, where the intra-cell and inter-cell signals are treated as interference. This yields (see, e.g., \cite{Tse05fundamentalsof})
\begin{align}
    &\text{KPI}(\mathbf{P}_0, \boldsymbol{\alpha}, \mathbf{H}_{\tau})= \sum_{c=1}^{N_C}\sum_{u=1}^{N_U} \log_2\det\bigg(\mathbf{I}_{N_R}\nonumber \\&+10^{\frac{P^{\mathrm{TX}}_{c,u}}{10}}\boldsymbol{\Gamma}_{c,u}^{-1}\mathbf{H}_{\tau,c,u,c}\mathbf{H}_{\tau,c,u,c}^{\sf H}\bigg)\quad [\text{bit/s/Hz}],\label{eq: mimo kpi}
\end{align}
where $\mathbf{I}_{N_R}$ is the $N_R\times N_R$ identity matrix, and $\boldsymbol{\Gamma}_{c,u}$ is the noise-plus-interference covariance matrix for the transmission of UE $u$  towards the serving BS in cell $c$, i.e.,
\begin{align}
    \boldsymbol{\Gamma}_{c,u} &= 10^{\frac{\sigma_z^2}{10}}\mathbf{I}_{N_R}+\sum_{j=1, j\neq u}^{N_U}10^{\frac{P^{\mathrm{TX}}_{c,j}}{10}}\mathbf{H}_{\tau,c,j,c}\mathbf{H}^{\sf H}_{\tau,c,j,c}\nonumber \\&+\sum_{c'=1,c'\neq c}^{N_C}\sum_{u=1}^{N_U}10^{\frac{P^{\mathrm{TX}}_{c',u}}{10}}\mathbf{H}_{\tau,c,u,c'}\mathbf{H}_{\tau,c,u,c'}^{\sf H},\label{eq: noise and interference}
\end{align}
with $\sigma_z^2$ as the channel noise power in logarithmic scale. Note that the transmitted powers $P^{\mathrm{TX}}_{c,j}$ from each $j$th UE in any cell $c$ are also measured in logarithmic scale.

\begin{figure}[t]

  \centering
  \centerline{\includegraphics[scale=0.4]{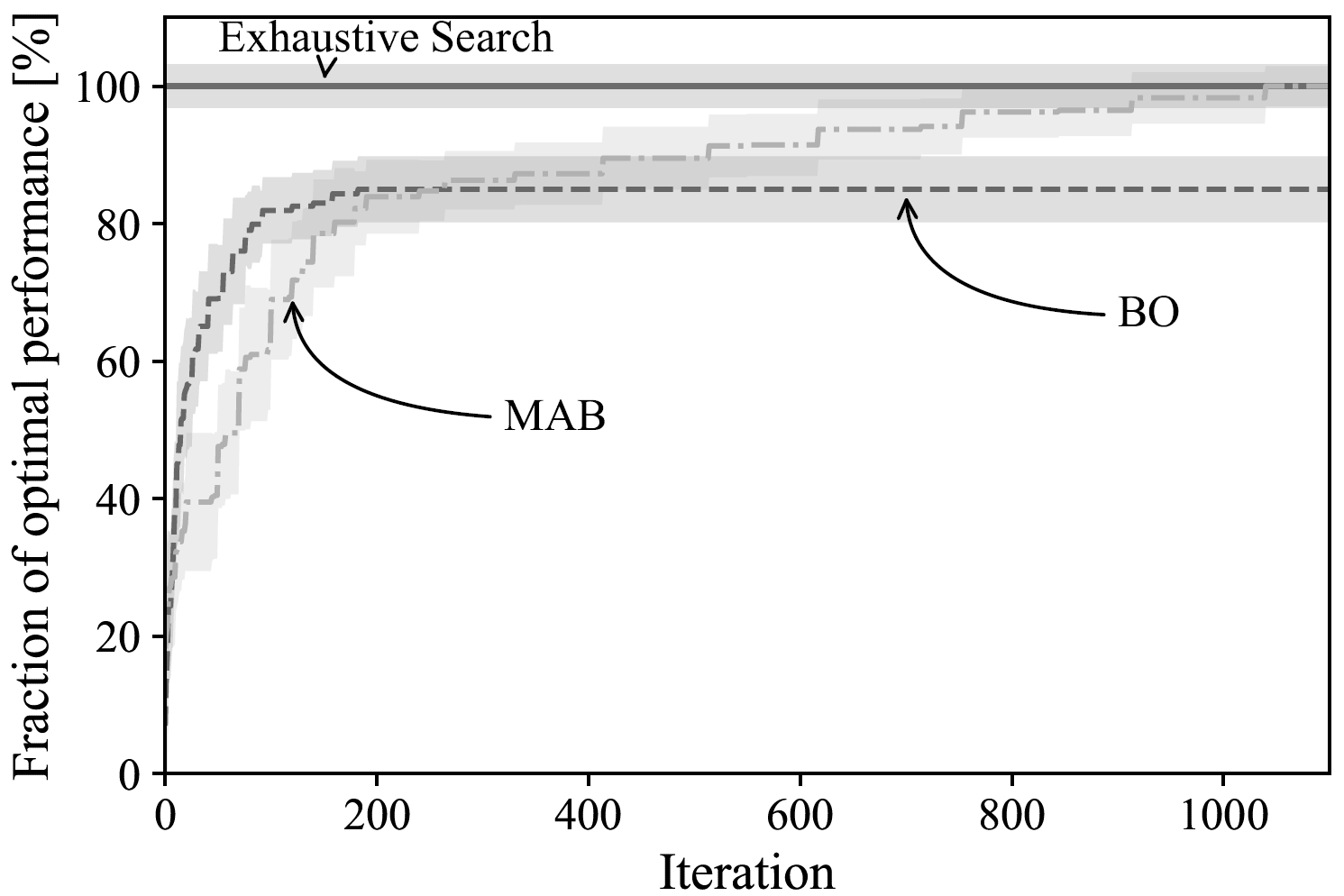}}
  \caption{Fraction of the optimal KPI \eqref{eq: mimo kpi} (compared to exhaustive search) obtained by BO and MAB optimizers for a multi-cell system as a function of the number of iterations of the optimization algorithms.}
  \label{fig: base}
%
\end{figure}

The joint distribution $\mathrm{P}_{\mathbf{H}_{\tau}}$ of the channel matrices $\mathbf{H}_{\tau}=\{\mathbf{H}_{\tau,c,u,c'}\}_{c=1,u=1,c'=1}^{N_C,N_U,N_C}$ depends on the wrap-around distance $\{d_{c,u,c'}\}$ between the $u$th UE in cell $c$ and the BS in cell $c'$ for $u=1,...,N_U$ and $c,c'=1,...,N_C$; on the receiver antennas height $h_{BS}$ relative to the UEs' height; on the power of shadow fading $\sigma_{SF}^2$; and on the carrier frequency $f_c$. Specifically, we model the $N_R\times N_T$ channel between UE $u$ in cell $c$ and the BS in cell $c'$ as\begin{align}
    \mathbf{H}_{\tau,c,u,c'}=10^{\frac{-\mathrm{PL}_{\tau,c,u,c'}}{20}}\beta_{\tau,c,u,c'}\mathbf{G}_{\tau,c,u,c'}, \label{eq: channel matrix}
\end{align}
where the distribution of the $N_R \times N_T$ random matrix $\mathbf{G}_{\tau,c,u,c'}$ and of the coefficient $\beta_{\tau,c,u,c'}$ depend on whether UE $u$ in cell $c$ is in non-line-of-sight (NLOS), or line-of-sight (LOS). With respect to BS $c'$, the LOS probability for each UE $u$ in cell $c$ is computed according to Table 7.4.2-1 in 3GPP TR 38.901 as
\begin{align}
    \mathrm{Pr}_{LOS,\tau,c,u,c'}=\begin{cases}1\\
    d_{\tau,c,u,c'}\leq d_{min},\\
    \frac{18}{d_{\tau,c,u,c'}}+\exp\Big(-\frac{d_{\tau,c,u,c'}}{36}\Big)\Big(1-\frac{18}{d_{\tau,c,u,c'}}\Big)\\d_{\tau,c,u,c'}>d_{min},\label{eq: line of sight prob}
\end{cases}
\end{align}
where $d_{min}$ is set to 18 m. The slow fading variable $\beta_{\tau,c,u,c'}$ is log-normal distributed with standard deviations $\sigma_{LOS,\tau}$ or $\sigma_{NLOS,\tau}$ with respective probabilities $\mathrm{Pr}_{LOS,\tau,c,u,c'}$ and  $1-\mathrm{Pr}_{LOS,\tau,c,u,c'}$; and the matrix $\mathbf{G}_{\tau,c,u,c'}$ is either Ricean or Rayleigh distributed with respective probabilities $\mathrm{Pr}_{LOS,\tau,c,u,c'}$ and  $1-\mathrm{Pr}_{LOS,\tau,c,u,c'}$. Furthermore, the pathloss $\mathrm{PL}_{c,u,c'}$ for LOS and NLOS, which are used in \eqref{eq: PUSCH power}, are obtained from the urban microcellular (UMi) street canyon pathloss model in Table 7.4.1-1 of 3GPP TR 38.901 as\begin{align}
    &\mathrm{PL}_{LOS,c,u,c'}=32.4+21\log_{10}(d'_{c,u,c'})+20\log_{10}(f_c), \nonumber \\&
    \mathrm{PL}_{NLOS,c,u,c'}=\max\Big(\mathrm{PL}_{LOS,c,u,c'}, 35.3\log_{10}(d'_{c,u,c'})\nonumber \\&+22.4+21.3\log_{10}(f_c)-0.3(h_{UE}-1.5)\Big), \label{eq: pathloss}
\end{align}
respectively, where $d'_{\tau,c,k}$ is the distance between UEs and receiver antennas in the wrap-around model. The parameter $\text{CL}_{u,c}$ in \eqref{eq: PUSCH power} is fixed to 0 dB in accordance to Table 7.2.1-1 in 3GPP TS 38.213.

We focus on the optimization of a single pair $(P_0,\alpha)$ of OLPC parameters shared across three cells. This relatively simple setting allows us to maximize function \eqref{eq: mimo kpi} exactly through exhaustive search, providing a useful benchmark for the considered approximate optimization strategies. 

\begin{figure}[t]

  \centering
	
  \includegraphics[scale=0.38]{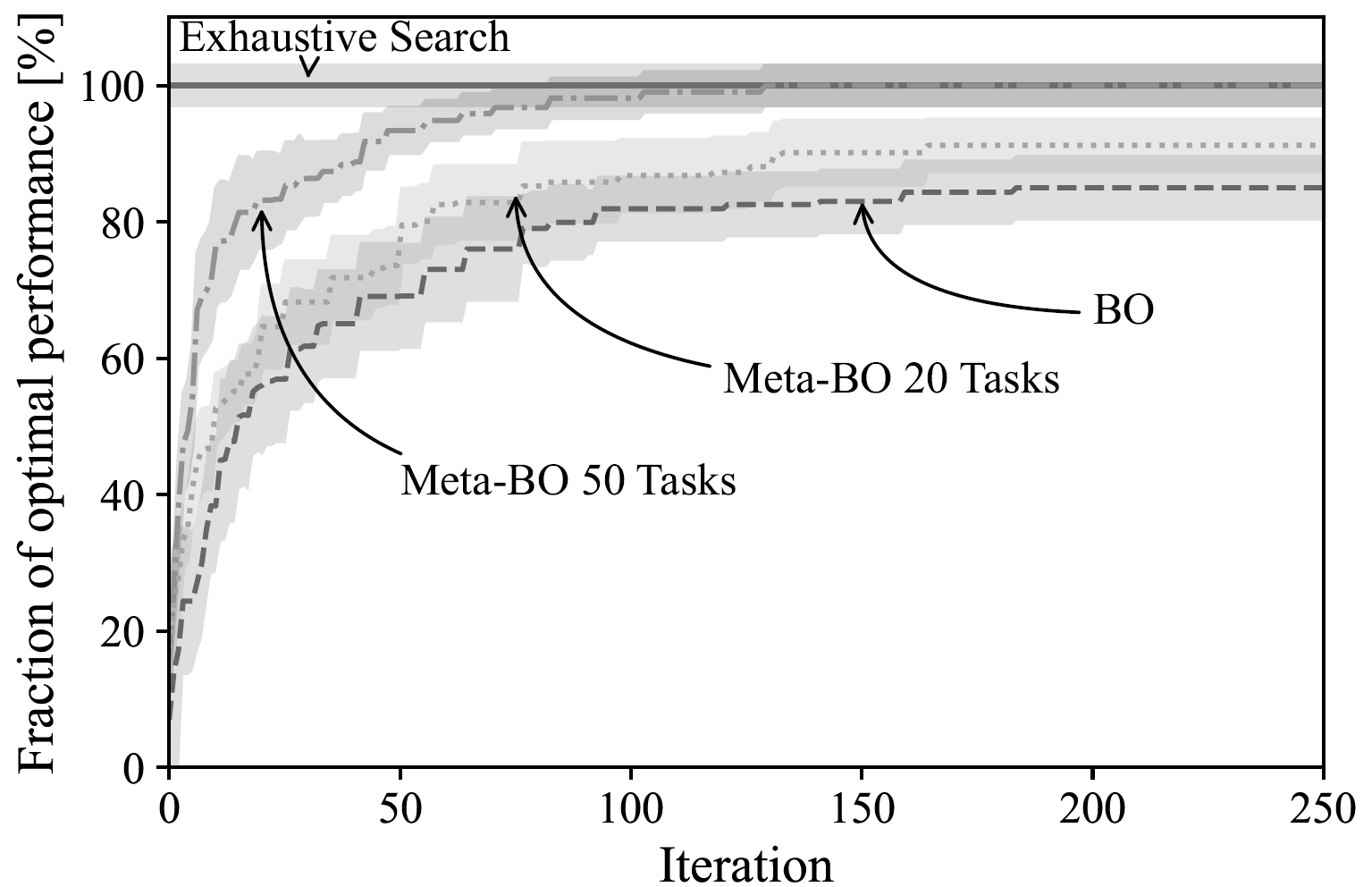}
  \includegraphics[scale=0.38]{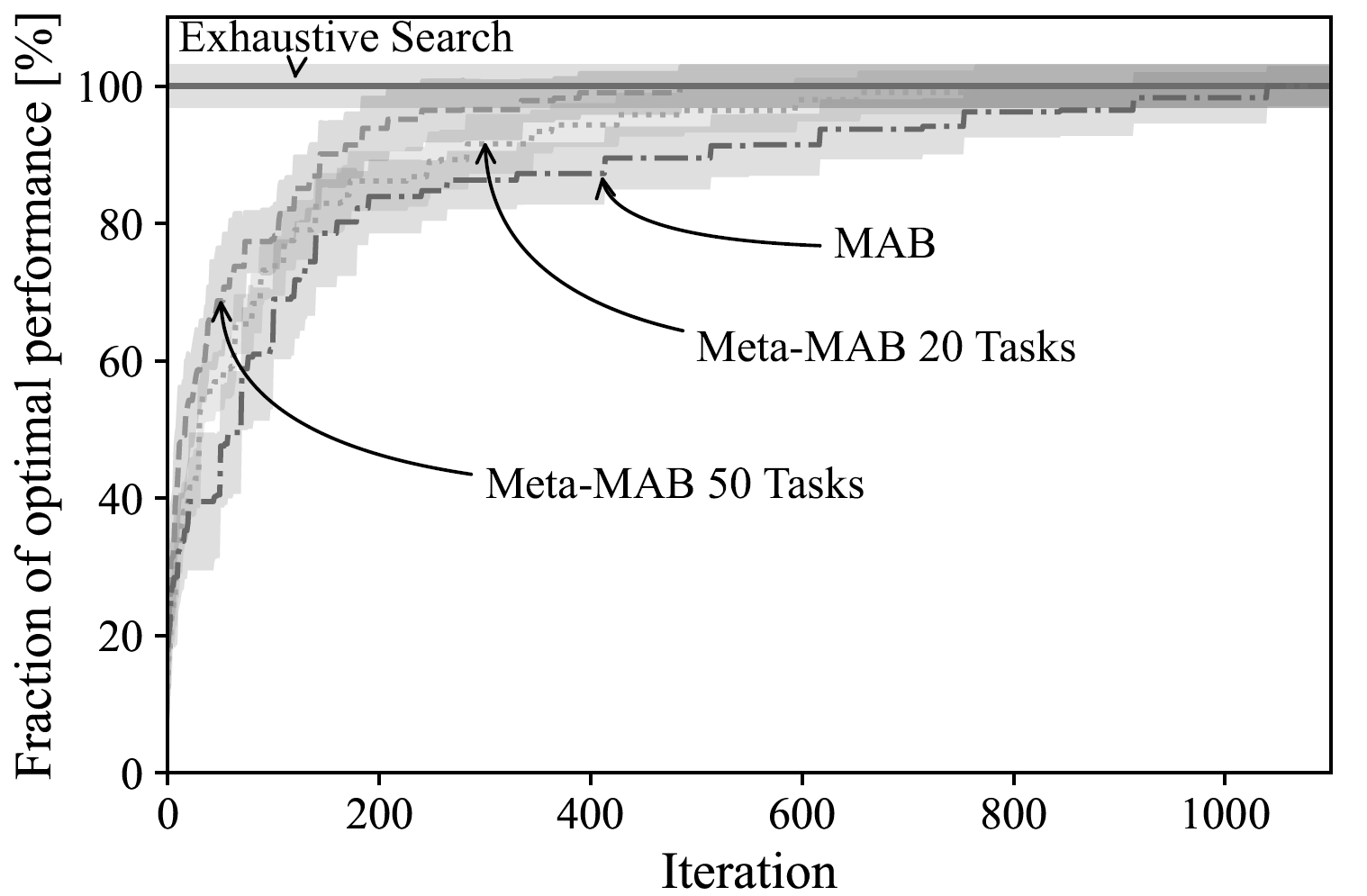}
  \caption{Fraction of the optimal KPI \eqref{eq: mimo kpi} (compared to exhaustive search) obtained by meta-BO (top) and meta-MAB (bottom) optimizers for a multi-cell system as a function of the number of iterations of the optimization algorithms.}
  \label{fig: meta}
%
\end{figure}

We fix the number of antennas to $N_R=16$ and $N_T=4$, the number of UEs to $N_U=10$ in each cell, the carrier frequency to $f_c=3.5$ GHz, the size of the CSI dataset for each configuration $\tau$ is set to $S_{\tau}=100$ samples, and the maximum transmit power is $P_{MAX,u}=23$ dBm for all UEs. 

For each configuration, the location of the UEs is fixed, and obtained by drawing distances $d_{c,u,c}$  to a serving BS uniformly in the interval $[18, 200]$ meters. As specified in UMi street canyon, the receiver height is $h_{BS}=15$ meters, the shadow fading standard deviations are set to 4 dB and 7.82 dB.
In accordance with Table 7.7.2-4 in 3GPP TR 38.901, Rayleigh fading variance is -13.5 dB for NLOS links, while Rice fading with mean -0.2 dB and variance -13.5 dB affects LOS UEs. The  noise power is set to $\sigma_z^2=-121.38$ dB.

\subsection{Conventional Bayesian and Bandit Optimization}\label{ssec: baselearner}

First, we evaluate the average KPI function \eqref{eq: noisy observation} using \eqref{eq: mimo kpi} in the full $(P_0, \alpha)$ solution space, where the KPI is averaged over 20 realizations of the dataset $D_{\tau}$ with the same configuration $\tau$. Fig. \ref{fig: exhaustive search} shows that the optimization target is multimodal, and hence generally computational challenging for traditional local search
algorithms.

\begin{figure}[t]

  \centering
	
  \includegraphics[scale=0.38]{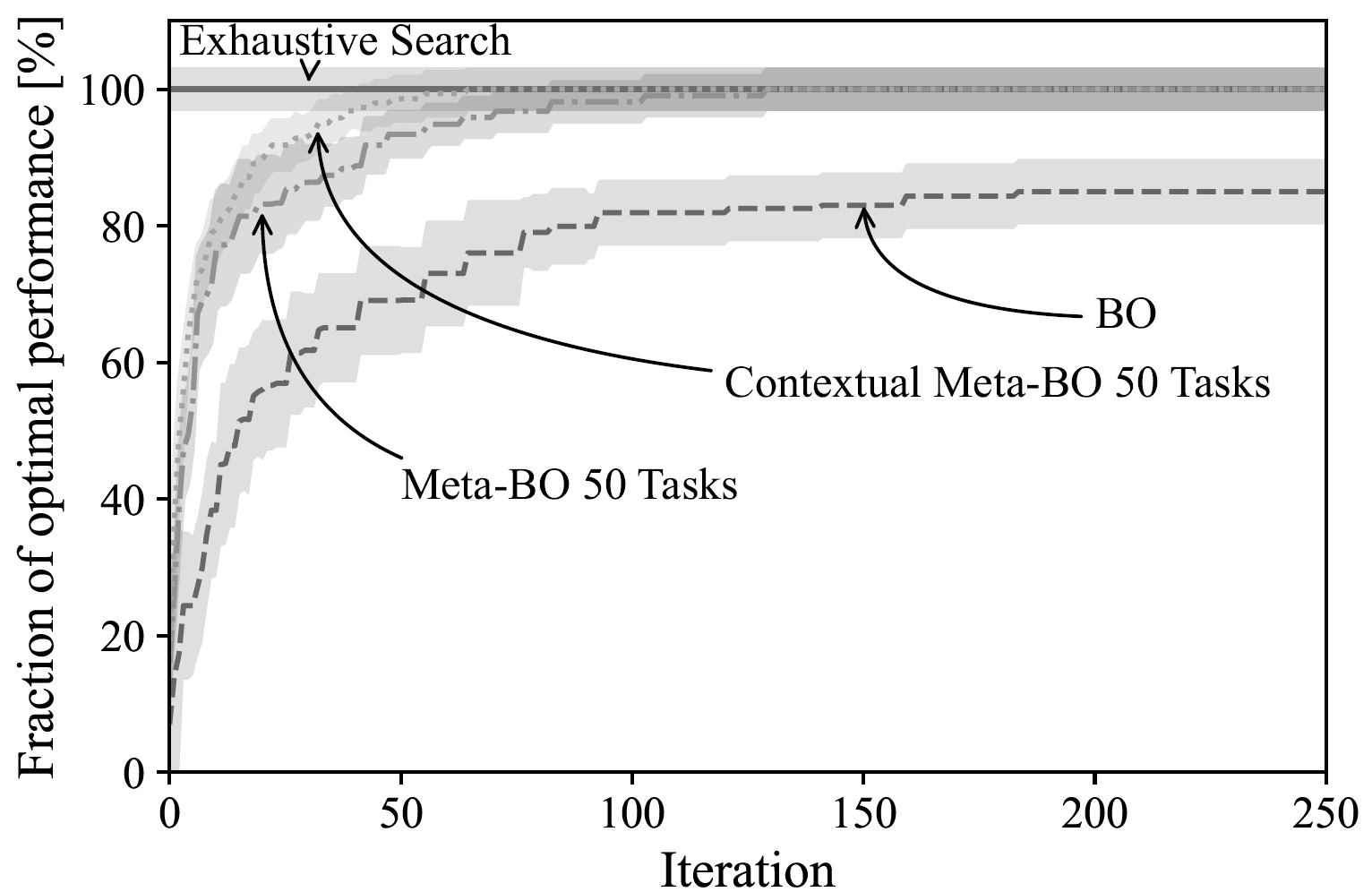}
  \includegraphics[scale=0.33]{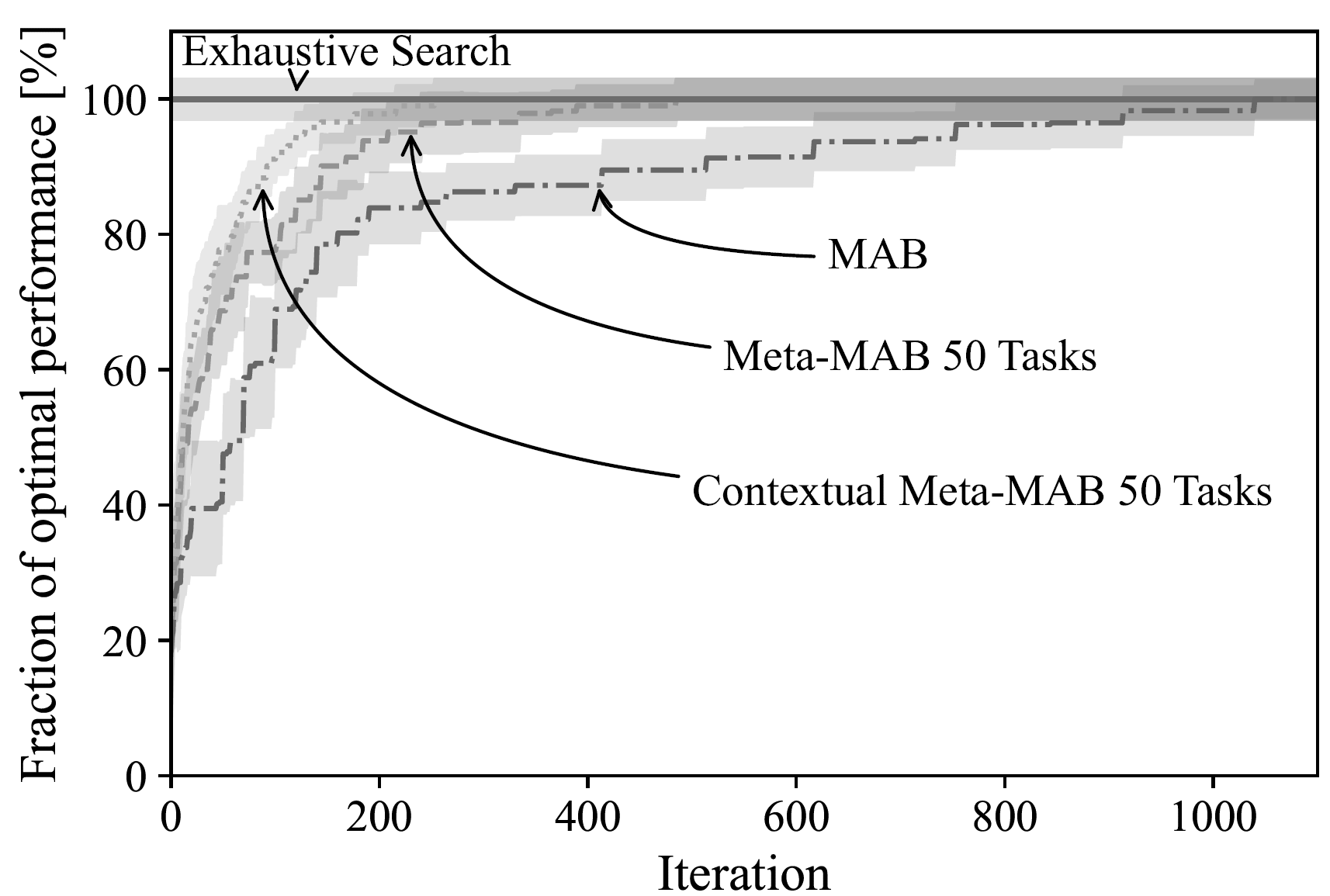}
  \caption{Fraction of the optimal KPI \eqref{eq: mimo kpi} (compared to exhaustive search) obtained by contextual meta-BO and vanilla meta-BO (top), as wel as by contextual meta-MAB and vanilla meta-MAB (bottom) for a multi-cell system as a function of the number of iterations of the optimization algorithms.}
  \label{fig: cont_meta}
%
\end{figure}

We now compare the performance of BO and bandit optimization on a single configuration $\tau$, with the performance averaged over 10 realizations and over 100 CSI datasets for each realization. We plot the KPI value normalized by the optimal value obtained via exhaustive search. The kernels for BO and bandit optimization are selected as \textit{Radial Basis Function kernels} (RBF) with bandwidth tuned to be 0.76 prior to the optimization, and we set parameter $\omega=0.3$ throughout the experiments for MAB via grid search. BO is seen to outperform bandit optimization for the first several iterations. At later iterations, the performance is limited by the  inherent bias of BO due to the continuous model used to approximate  optimization in a discrete space. This causes bandit optimization, which operates directly on a discrete space, to outperform BO when the number of iterations is sufficiently large, attaining the performance of exhaustive search.

\subsection{Bayesian and Bandit Meta-Optimization}

Having observed the relative inefficiency of BO and MAB in terms of number of iterations in Fig. \ref{fig: base}, we now evaluate the performance of Bayesian meta-optimization (Algorithm \ref{table: BMO}) and bandit meta-optimization (Algorithm \ref{table: meta mab}). We refer to these schemes for short as \emph{meta-BO} and \emph{meta-MAB}, respectively. Both the parametric mean function $\mu_{\boldsymbol{\theta}}(\cdot)$ and function $\psi_{\boldsymbol{\theta}}(\cdot)$ for kernels \eqref{eq: parametric function} are instantiated as  fully-connected neural networks with 3 layers with each 32 neurons. Setting the number of meta-training configurations to $N=50$, and the number of collected data pairs to $T_n=10$, Fig. \ref{fig: meta} shows the fraction of the optimal KPI  for both meta-optimization strategies. 

It is observed that meta-learning accelerates the convergence for both BO and MAB. For example, meta-MAB with 50 tasks can achieve a 90\% fraction of the optimal performance after around 175 iterations, while conventional MAB would require around 510 iterations. However, as the number of iterations increases, the gain of meta-MAB over MAB vanishes, since MAB is already able to achieve the performance of exhaustive search given its direct  optimization in the discrete space.

In this regard, BO stands to gain more from the implementation of meta-learning, since, as seen in Fig. \ref{fig: base}, the performance of BO is limited by the bias caused by the optimization over a continuous space as the number of iterations increase. For instance, with data from 50 tasks, meta-BO can achieve a 90\% fraction of the optimal performance already at 50 iterations, while conventional BO would not be able to obtain this performance level. More generally, meta-BO with 50 tasks can achieve any desired performance level in less than around 150 iterations. This indicates that optimizing the kernel via meta-learning can fully compensate for the bias caused by the fact that BO addresses the optimization problem in a continuous design space.  

Overall, while, without meta-learning, MAB is preferable over BO if the goal is achieving high-quality solutions, as long as data from a sufficiently large number of tasks is available, meta-BO becomes significantly advantageous. For the example at hand, as mentioned, a 90\% performance level is obtained with meta-MAB with around 175 iterations. while meta-BO requires only 50 iterations.

\begin{figure}[t]

  \centering
	
  \includegraphics[scale=0.335]{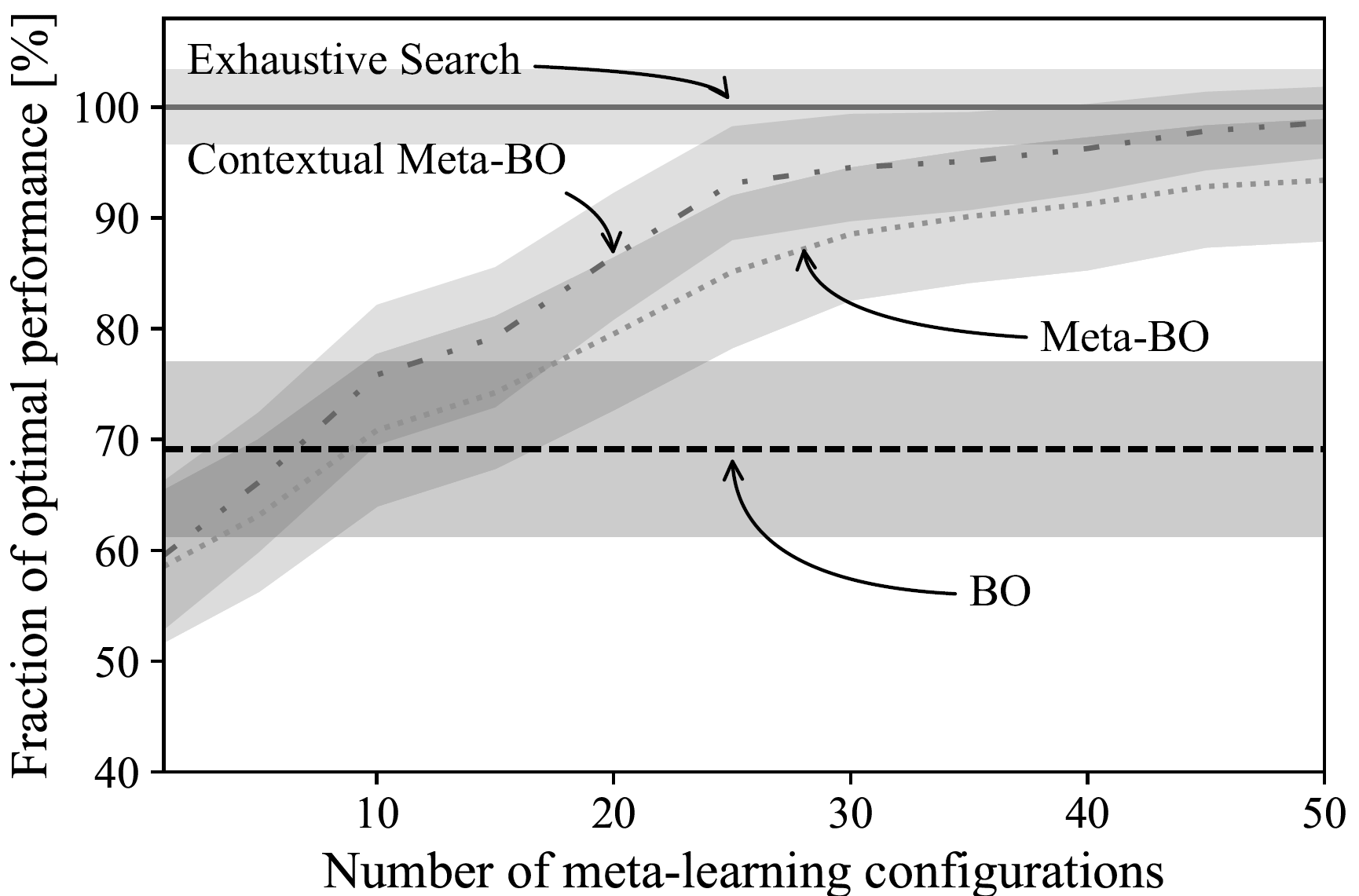}
  \includegraphics[scale=0.335]{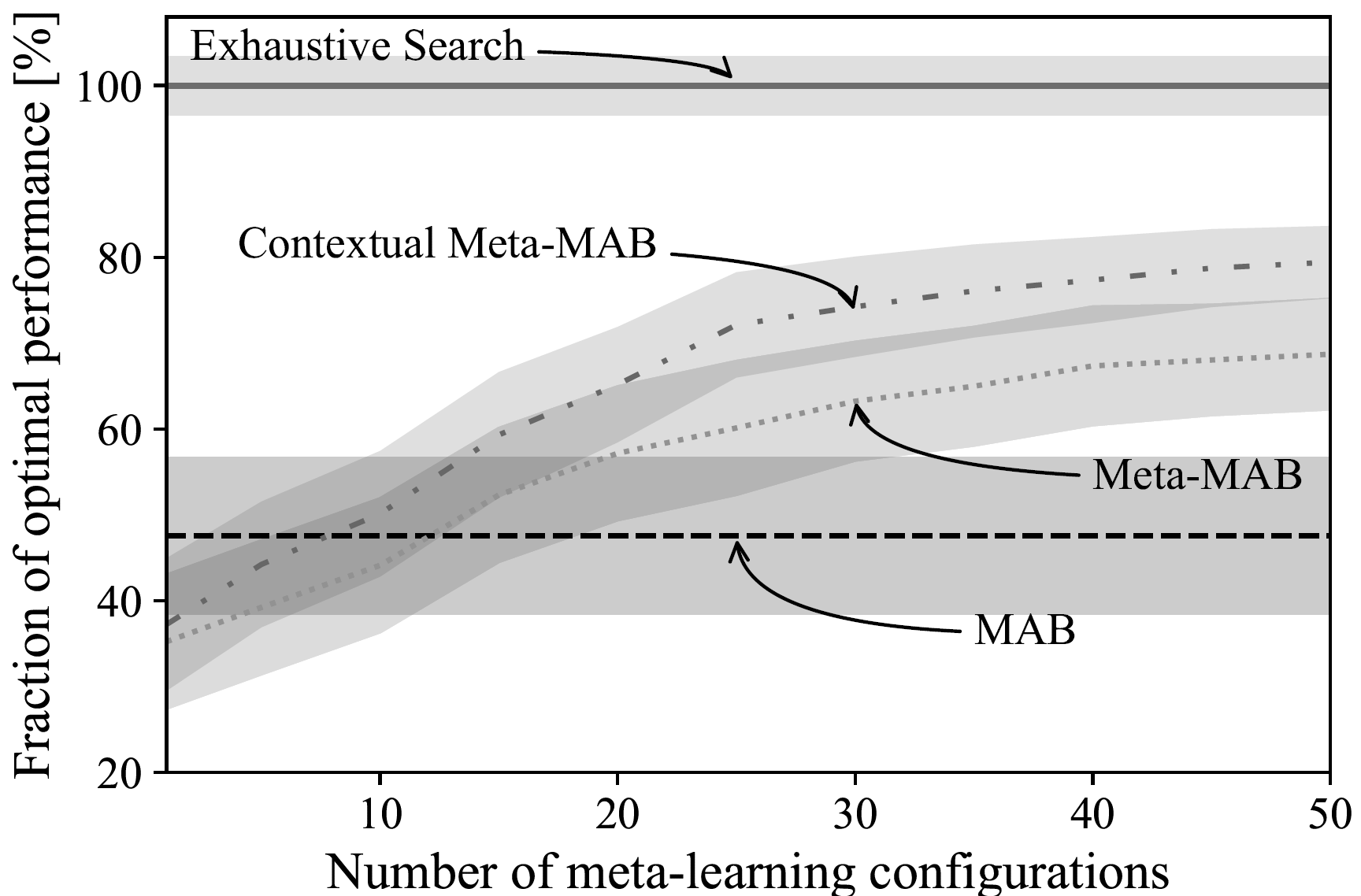}
  \caption{Fraction of the optimal KPI \eqref{eq: mimo kpi} (compared to exhaustive search) obtained by contextual meta-BO and vanilla meta-BO (top), as wel as by contextual meta-MAB and vanilla meta-MAB (bottom) for a multi-cell system as a function of the number of available meta-training configurations.}
  \label{fig: tasks}
%
\end{figure}

\subsection{Contextual Bayesian and Bandit Meta-Optimization}

We now investigate the performance of contextual Bayesian meta-optimization (Sec. \ref{ssec: cbmo}) and contextual Bandit meta-optimization (Sec. \ref{ssec: cmmab}), which we refer for short as \emph{contextual meta-BO} and \emph{contextual meta-MAB}, respectively. We are interested in addressing the potential benefits as compared to vanilla meta-BO and meta-MAB. In order to obtain the interference graph, the threshold ratio $d_{ji}/d_i$ is set to 1.8; and the rank of the parameter matrices $\mathbf{V}_1, \mathbf{V}_2$ is set to $r=14$ for both algorithms. Both values are obtained via a coarse grid search. The number of meta-training tasks is set to $N=50$. 

Fig. \ref{fig: cont_meta} demonstrates the fraction of optimal KPI for both context-based strategies as compared to the vanilla counterpart solutions. The results validate the capacity of the proposed contextual meta-learning methods to extract useful information from the network topology for the given configuration, achieving faster convergence for both Meta-BO and Meta-MAB.

\begin{figure}[t]

  \centering
	
  \includegraphics[scale=0.48]{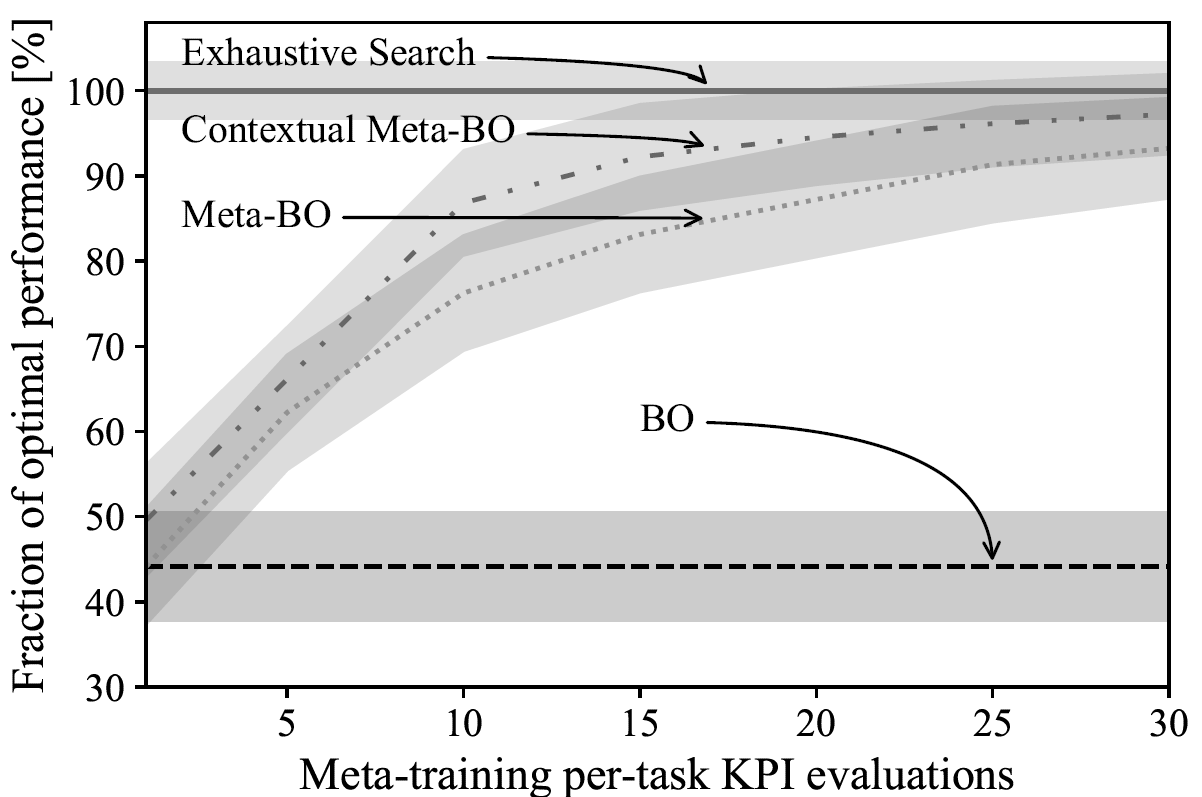}
  \includegraphics[scale=0.48]{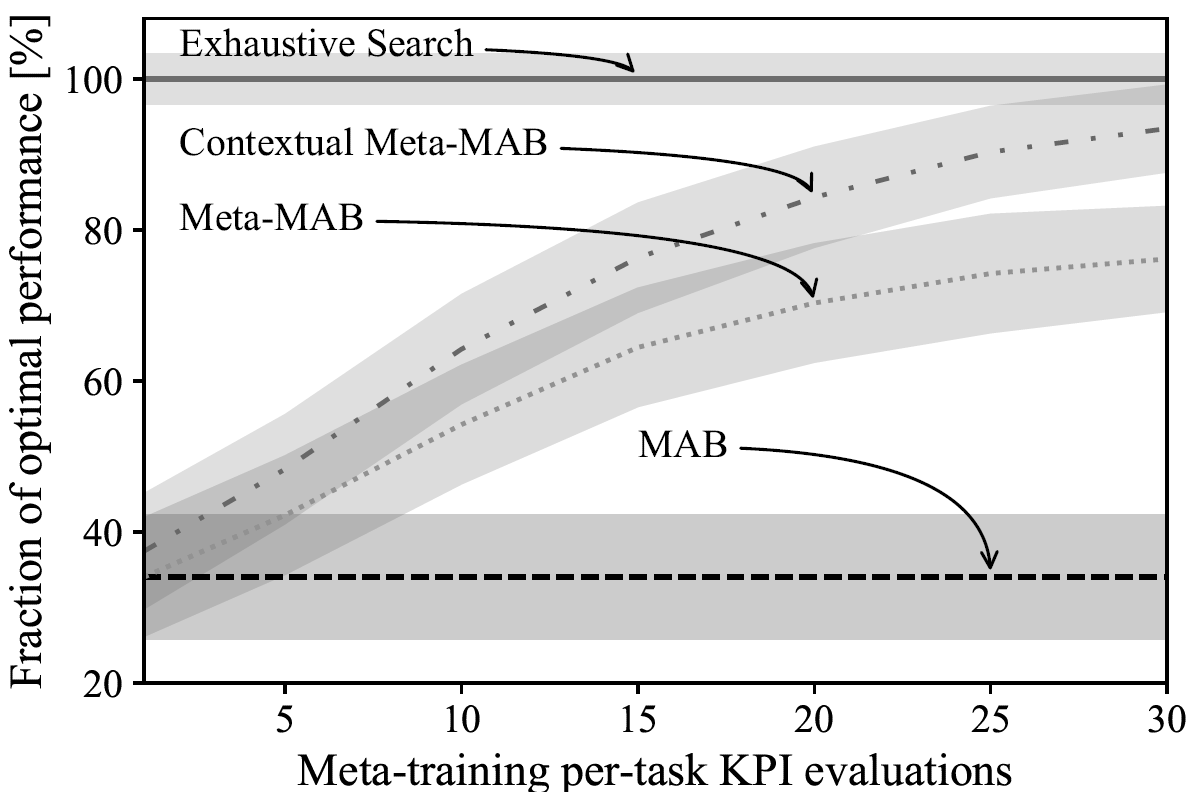}
  \caption{Fraction of the optimal KPI \eqref{eq: mimo kpi} (compared to exhaustive search) obtained by contextual meta-BO and vanilla meta-BO (top), as wel as by contextual meta-MAB and vanilla meta-MAB (bottom) for a multi-cell system as a function of the number of available KPI evaluations $T_n$ per-meta-training task.}
  \label{fig: Tn}
%
\end{figure}

We elaborate on the impact of the number $N$ of meta-training tasks in Fig. \ref{fig: tasks}, which shows the fraction of optimal KPI obtained at the 50th iteration. It is observed that  a number of meta-training tasks equal to $N=10$ for meta-BO and $N=12$ for meta-MAB is sufficient to ensure that vanilla meta-BO and meta-MAB optimizers can transfer useful information from the meta-training configurations to the new configurations to speed up optimization as compared to BO and MAB, respectively. Furthermore, contextual meta-BO and contextual meta-MAB can further decrease the number of required meta-training configurations.

Finally, we address the impact of the number $T_n$ of per-task KPI evaluations available in the meta-training data. We evaluate the fraction of the optimal KPI obtained at the 20th iteration, and set $N=50$ tasks. In Fig. \ref{fig: Tn}, we observe that meta-BO and meta-MAB, as well as their contextual versions, can significantly enhance the performance of vanilla BO and MAB with as few as $T_n=20$ KPI evaluations per task. Concretely, while vanilla BO obtains a fraction around 40$\%$ of the optimal performance, with $T_n=20$,  contextual BO achieves more than 90$\%$ of this fraction, providing a 10$\%$ gain over meta-BO. Similarly, while vanilla MAB obtains 30$\%$  of the optimal performance, with $T_n=20$,  meta-MAB obtains a 70$\%$ fraction, and contextual MAB an 80$\%$ fraction.

\section{Conclusions}

Modern cellular networks require complex resource allocation procedures that can only leverage limited access to KPI evaluations for different candidate resource-allocation parameters. While data collection  for the current network deployment of interest is challenging, a network operator has typically access to data from related, but distinct, deployments. This paper has proposed to transfer knowledge from such historical or simulated deployments via an offline meta-learning phased with the aim of learning how to optimize on new deployments. As such, the proposed meta-learning approach can be integrated with digital twin platform providing simulated data \cite{Ruah2022}. We have specifically focused on BO and MAB optimizers, with the former natively operating on a continuous optimization domain and the latter on a discrete domain. Furthermore, we have proposed novel BO and MAB-based optimizers that can integrate contextual information in the form of interference graphs into the resource-allocation optimization. Experimental results have validated the efficiency gains of meta-learning and contextual meta-learning.

Future work may address online meta-learning techniques that successively improve the efficiency of resource allocation as data from more deployments is (see \cite{park2020learning} for a related application to demodulation and \cite{marini2022continual} to drone trajectory optimization). Furthermore, it would be interesting to investigate the application to larger-scale problems involving real-world data; the extension to multi-objective problems \cite{sergey2022cellfree}; and the interplay with digital twin platforms for the management of wireless systems \cite{Ruah2022}.

\ifCLASSOPTIONcaptionsoff
  \newpage
\fi

\bibliographystyle{IEEEtran}
\bibliography{refer}





\end{document}